\documentclass[aps,prd,reprint,superscriptaddress,longbibliography,floatfix]{revtex4-2}

\usepackage[T1]{fontenc}   
\usepackage[utf8]{inputenc} %
\usepackage{lmodern}        

\usepackage{orcidlink}
\usepackage{graphicx}
\usepackage{dcolumn}
\usepackage{bm}
\usepackage{hyperref}
\hypersetup{
  colorlinks=true,
  allcolors=blue!60!black, 
  pdfborder={0 0 0},
}

\usepackage{booktabs}
\setcitestyle{numbers,square,comma,sort&compress}

\usepackage{xcolor}

\begin{document}

\title{Simulation-Based Inference for Direction Reconstruction of Ultra-High-Energy\\ Cosmic Rays with Radio Arrays}


\author{Oscar Macias\,\orcidlink{0000-0001-8867-2693}}
\email{macias@sfsu.edu}
\affiliation{Department of Physics and Astronomy, San Francisco State University, San Francisco, CA 94132, USA}
\affiliation{GRAPPA -- Gravitational and Astroparticle Physics Amsterdam, University of Amsterdam, Science Park 904, 1098 XH Amsterdam, The Netherlands}

\author{Zachary Mason\,\orcidlink{0009-0003-2423-6667}}
\email{zmason2@sfsu.edu}
\affiliation{Department of Physics and Astronomy, San Francisco State University, San Francisco, CA 94132, USA}


\author{Matthew Ho\,\orcidlink{0000-0003-3207-8868}}
\affiliation{Department of Astronomy, Columbia University, New York, NY 10027, USA}
\author{Ars\`ene Ferri\`ere}
\affiliation{Université Paris-Saclay, CEA, List, F-91120, Palaiseau, France}
\affiliation{Sorbonne Université, CNRS, Laboratoire de Physique Nucléaire et des Hautes Energies
(LPNHE), 4 Pl. Jussieu, Paris, 75005, France}

\author{Aur\'elien Benoit-L\'evy}
\affiliation{Université Paris-Saclay, CEA, List, F-91120, Palaiseau, France}

\author{Mat\'ias Tueros \orcidlink{0000-0003-1570-1419}}
\affiliation{IFLP - CCT La Plata - CONICET, Diag 113 y 63, La Plata (1900), Argentina} \affiliation{Depto. de Física, Fac. de Cs. Ex., Universidad Nacional de La Plata,  Casilla de Correo 67, La Plata (1900), Argentina}


\begin{abstract}

Ultra-high-energy cosmic-ray (UHECR) observatories require unbiased direction reconstruction to enable multi-messenger astronomy with sparse, nanosecond-scale radio pulses. Explicit likelihood methods often rely on simplified models, which may bias results and understate uncertainties. We introduce a simulation-based inference pipeline that couples a physics-informed graph neural network (GNN) to a normalizing-flow posterior within the \textit{Learning the Universe Implicit Likelihood Inference} framework. Each event is seeded by an analytic plane-wavefront fit; the GNN refines this estimate by learning spatiotemporal correlations among antenna signals, and its frozen embedding conditions an eight-block autoregressive flow that returns the full Bayesian posterior. Trained on about $8,000$ realistic UHECR air-shower simulations generated with the ZHAireS code, the posteriors are temperature-calibrated to meet empirical coverage targets. We demonstrate a sub-degree median angular resolution on test UHECR events, and find that the nominal 68\% highest-posterior-density contours capture $71\% \pm 2\%$ of true arrival directions, indicating a mildly conservative uncertainty calibration. This approach provides physically interpretable reconstructions, well-calibrated uncertainties, and rapid inference, making it ideally suited for upcoming experiments targeting highly inclined events, such as GRAND, AugerPrime Radio, and BEACON.

\end{abstract}
\maketitle

\section{Introduction}\label{sec:intro}

Ultra-high-energy cosmic rays (UHECRs) are the most energetic particles detected in nature~\cite{HIRES:1994ijd, TelescopeArray:2023sbd}, but their origins and propagation mechanisms remain poorly understood~\cite{Anchordoqui:2018qom, Ahlers:2015lln, AlvesBatista:2024czs}. Accurate detection and reconstruction of these rare events have significant implications for fundamental physics and astronomy~\cite{AlvesBatista:2019tlv, Bustamante:2023iyn, Guepin:2022qpl}. 
In particular, accurate arrival directions underpin inference for UHECRs (they set the shower-axis direction and, together with the core position, the event geometry used by downstream reconstructions), so mis-pointing biases the depth of shower maximum and the energy scale that feed composition studies~\cite{Decoene:2021ncf, Apel:2014usa, Schluter:2020tdz}. For ultra-high-energy neutrinos, precise pointing is essential for source association and for identifying Earth-skimming trajectories that arrive from below the horizon~\cite{Schroder:2016hrv}.

Next-generation observatories are rising to this challenge by focusing on radio detection techniques, which offer all-sky coverage and high-duty cycles~\cite{Huege:2016veh}. Large-scale projects such as the planned Giant Radio Array for Neutrino Detection (GRAND)~\cite{GRAND:2018iaj}, AugerPrime Radio~\cite{PierreAuger:2016qzd}, 
and BEACON~\cite{Southall:2022yil}
are designed to instrument vast areas with antenna networks to capture the nanosecond radio pulses from extensive air showers. These instruments are particularly sensitive to inclined (nearly horizontal) air showers, which produce broad radio footprints on the ground and are a prime channel for detecting Earth-skimming neutrinos and high-energy cosmic rays near the horizon. With their large apertures, radio arrays open a new window for ultra-high-energy neutrino detection. For UHECRs, radio provides high-duty-cycle measurements and can constrain shower development relevant to composition in non- to moderately-inclined geometries. Extending robust composition sensitivity to very inclined showers remains an active area of development and may require hybrid measurements.

The detection of UHECRs using radio arrays is now a well-established technique. Modern radio arrays---such as LOPES~\cite{LOPES:2005ipv}, CODALEMA~\cite{Ardouin:2005qe}, LOFAR~\cite{VanHaarlem:2013abc}, AERA~\cite{Huege:2019snr}, and TREND~\cite{Charrier:2018fle}; reconstruct the incoming direction of UHECRs by fitting a geometric wavefront model to the measured trigger times at known antenna positions.

A reconstruction pipeline has recently been proposed~\cite{Decoene:2021ncf} for GRAND that leverages the amplitude distribution of the radio signal in a physics-informed, multi-stage approach. The procedure begins with a plane wavefront (PWF) timing fit to obtain an initial estimate of the arrival direction. This initial estimate is subsequently refined by introducing a mild curvature represented by a spheroidal surface wavefront model. In the final step, the lateral distribution of the radio amplitude is introduced to constrain the shower core position and axis orientation precisely. This hierarchical approach yields robust reconstruction performance, especially for air showers at high zenith angles~\cite{Guelfand:2025goo}.

\begin{figure*}[t!]
    \centering
    \includegraphics[width=1.0\linewidth]{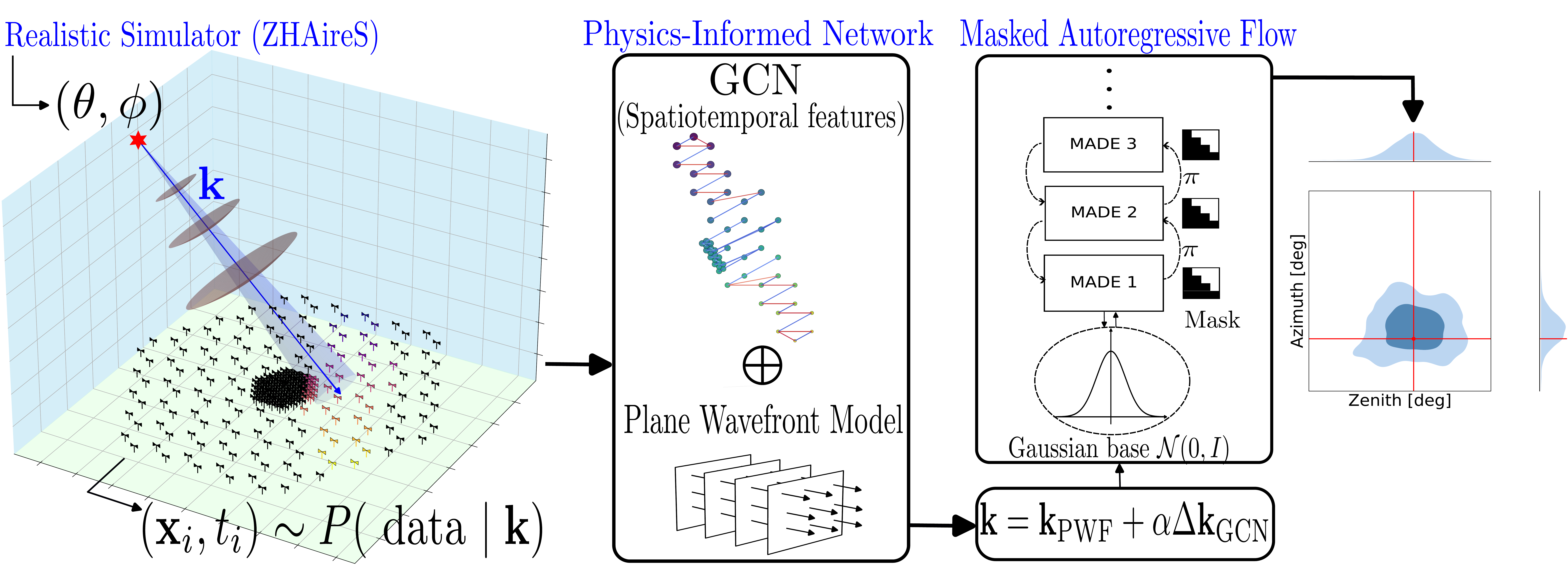}
    \caption{\textbf{End-to-end SBI pipeline within the \texttt{LtU-ILI} framework:} Antenna signals simulated with \texttt{ZHAireS} are compressed by a physics-informed Graph Convolutional Network (GCN) and plane-wavefront (PWF) model into a residual offset $\Delta\mathbf{k}$. Adding this residual to an analytic fit yields the three-dimensional arrival direction $\mathbf{k}$, forming an information bottleneck focused on geometry. A masked autoregressive flow transforms a Gaussian base distribution into the posterior $p(\mathbf{k}\mid\text{data})$. Each flow layer employs a Masked Autoencoder Distribution Estimator (MADE), with triangular masks enforcing causality, resulting in a strictly lower-triangular Jacobian that enables efficient autoregressive posterior density evaluation without explicit likelihood computations.
}
    \label{fig:illustration}
\end{figure*}

Although extremely promising, this method also exposes the limits of current reconstruction pipelines. Its successive steps (i.e., planar fitting, curvature correction, and amplitude weighting) rely on model-dependent assumptions that could introduce biases. Traditional \textit{explicit likelihood methods}, such as formulating a likelihood for observed arrival times given a trial geometry and exploring it via Markov Chain Monte Carlo, are conceptually appealing but practically prohibitive for large arrays and complex wavefront structures~\cite{Guelfand:2025goo}. Template‐matching schemes confront similar scalability bottlenecks; reproducing the full span of arrival directions, primary energies, and atmospheric states demands vast libraries of simulated air-shower footprints, whose storage and compute requirements grow steeply with each gain in angular or temporal resolution \citep{Desmet:2025ufy}.

Equally important, conventional reconstruction algorithms often fail to rigorously quantify uncertainties; while they may yield a best-fit arrival direction, recovering the full posterior distribution remains non-trivial and is frequently overlooked. The recent analytical PWF solution in Ref.~\cite{Ferriere:2025} addresses part of this gap by providing a closed-form direction estimator with error estimates for the planar approximation. Yet, a planar wavefront model can only capture so much. To fully exploit the rich information in the radio signal pattern, more flexible inference approaches are needed.

A promising route to overcome the limitations of explicit likelihood methods is \textit{simulation-based inference} (SBI)~\cite{Cranmer:2019eaq}, also known as implicit or likelihood-free inference. Rather than relying on an analytic likelihood function, SBI leverages high-fidelity simulations of the physics process to directly learn the relationship between model parameters (e.g. shower direction) and observables (antenna times and signals). In a Bayesian context, SBI methods aim to directly estimate the posterior distribution of parameters by training machine learning models on simulated data, thereby \textit{``learning''} an approximate likelihood function or posterior emulator (see Ref.~\cite{Miller:2020hua, Ho:2024whi} and references therein).

This approach is well suited to astroparticle problems like UHECR air showers, where trigger times at fixed stations can be modeled with simple parametric likelihoods under geometric wavefront models (e.g., plane or hyperbolic with Gaussian jitter)~\cite{Apel:2014usa, VanHaarlem:2013abc, Huege:2019snr}. However, a tractable closed-form likelihood for the \emph{full} radio data vector (multi-polarization waveforms across many stations, non-Gaussian and correlated noise, thresholds, and array-response systematics) is lacking. High-fidelity forward models such as CoREAS~\cite{Huege:2016veh} and ZHAireS~\cite{Alvarez-Muniz:2012} capture these effects only through simulation. Simulation-based inference leverages such simulators to perform Bayesian inference without prescribing an explicit analytic likelihood, thereby avoiding restrictive residual models while keeping assumptions transparent in the simulator and prior~\cite{Cranmer:2019eaq}.

In recent years, SBI techniques have gained traction across astrophysics and cosmology, often outperforming traditional analysis methods~\cite{Mishra-Sharma:2021oxe, Montel:2022fhv, Tucci:2023bag, Modi:2023llw, FrancoAbellan:2024tbj, Bhardwaj:2023xph, Karchev:2024zpu, Savchenko:2025jzs, Eckner:2025waa, List:2025qbx}. Examples range from inferring cosmological parameters from galaxy surveys to decoding gravitational wave signals, where neural density estimators have achieved faster and more accurate results than classical MCMC in high-dimensional parameter spaces (See also Ref.~\cite{Miller:2020hua} for a detailed discussion). These successes underscore the appeal of SBI; it provides fully Bayesian posteriors with uncertainty quantification for problems that are otherwise intractable with explicit likelihoods~\cite{Ho:2024whi}. Importantly, likelihood-free approaches can naturally handle complicated selection effects or ``cuts'' on the data by applying the same cuts to simulated data, a notable advantage when dealing with complex experimental setups. 

One such framework is \textit{Learning the Universe---Implicit Likelihood Inference (LtU-ILI)}~\cite{Ho:2024whi}. \texttt{LtU-ILI} provides an end-to-end SBI platform that incorporates adaptable neural encoders, inference strategies, and rigorous validation tools. A notable feature of \texttt{LtU-ILI} is its support for complex data representations through custom ``embedding'' networks---including convolutional networks for image-like data and graph neural networks (GNNs) for relational data.

This is particularly pertinent for radio detector arrays such as GRAND, where the spatial configuration of antennas and their trigger times can be naturally modeled as a graph (with antennas as nodes and edges connecting neighbor detectors). Such a representation enables graph neural networks to effectively capture spatiotemporal correlations in the developing air-shower front. Moreover, embedded in a Bayesian workflow, \texttt{LtU-ILI} audits its posteriors with coverage tests to verify statistical calibration \citep{Ho:2024whi}. The result is an inference pipeline that remains computationally tractable while preserving clear physical meaning.

In this article, we present a novel UHECR direction reconstruction pipeline that combines a physics-informed neural network, integrating a graph neural network with a PWF model, to process antenna positions and trigger-time data generated from thousands of realistic ZHAireS simulations. This physics-guided embedding initializes a normalizing-flow model within the LtU-ILI framework, which is trained on full simulation data to yield posterior cones in the sky. The hybrid approach unites the interpretability of analytical models with the flexibility of deep learning and the rigor of simulation-based inference. It delivers statistically calibrated posteriors, a transparent uncertainty budget for multimessenger follow-up, and scalability to next-generation detector arrays. 

The remainder of this paper is organized as follows. Section~\ref{Sec:pipeline} describes the simulation dataset, the baseline PWF reconstruction algorithm, the design of the graph neural network, and the LtU-ILI inference setup. In Section~\ref{sec:Results} we present the performance of the combined pipeline on simulated events, highlighting the gains in angular resolution for inclined showers and the calibration of the uncertainty estimates. We discuss the implications for upcoming experiments, such as GRAND, and potential extensions of this method for additional UHE cosmic messenger properties in Section~\ref{sec:discussion}. Finally, Section~\ref{sec:conclusion} provides our conclusions and outlook.

The code used in this work is publicly available (see Data and Code Availability in Sec.~\ref{sec:data-availability}).

\section{Simulation-based inference pipeline}\label{Sec:pipeline}

Figure~\ref{fig:illustration} sketches our simulation-based inference pipeline at a glance. In particular, we feed raw antenna signals from \texttt{ZHAireS} into a physics-guided graph network that blends a plane-wavefront seed with learned residuals, yielding a compact geometric vector $\Delta\mathbf{k}$.  This three-dimensional summary becomes the input to an eight-block masked autoregressive flow that maps a Gaussian base to the full posterior $p(\mathbf{k}\!\mid\!\text{data})$.  The sections below unpack each stage. Appendix~\ref{appx:A} and \ref{appx:B-Flow-architecture} supply architecture tables, training schedules, and reproducibility notes.

\subsection{Physical Modeling  of Extensive Air Showers}\label{subsec:physofairshowers}

When an UHECR primary strikes the atmosphere, it initiates a cascade of secondary particles known as an extensive air shower (EAS). The primary collision produces a multitude of daughter particles that branch into a hadronic component (secondary mesons like pions and their decay products) and an electromagnetic component (e.g., photons from $\pi^0$ decay feeding $e^+e^-$ cascades)~\cite{Huege:2016veh}. This process continues until the primary’s energy is dissipated, yielding a shower of billions of particles spread over kilometers by the time it reaches ground level. Accurately modeling these cascades is essential, as the spatial, temporal, and energy distributions of the secondaries encode the nature of the primary particle~\cite{Alvarez-Muniz:2012}. In particular, the arrival direction of the primary can be inferred from the pattern of particle arrival times or atmospheric emissions, providing critical clues to the UHECR's origin~\cite{Decoene:2021ncf, Guelfand:2025goo, Ferriere:2025}.

Our work focuses on radio-detection of air showers, a technique that has matured over the past two decades. As the EAS evolves, its charged component experiences geomagnetic deflection and charge separation in the Earth’s magnetic field, driving transverse currents that emit coherent radio pulses. Additionally, a negative charge excess builds up in the shower front (due to electrons knocking out ambient electrons from air molecules), leading to Askaryan radiation~\cite{Huege:2016veh, Schroder:2016hrv}. The total radio signal is a superposition of these geomagnetic and Askaryan contributions. Modern experiments like LOFAR~\cite{VanHaarlem:2013abc}, AERA~\cite{Huege:2019snr}, and others have demonstrated that radio arrays can measure UHECR showers with impressive precision, rivaling conventional particle detectors in reconstructing the shower’s arrival direction and even its energy and shower maximum ($X_{\rm max}$). Radio detection thus provides a robust method to observe UHECR events over extensive areas and under most weather conditions (with the important exception of thunderstorms periods). This is achieved by recording the nanosecond-scale electromagnetic pulses generated by air showers using \emph{broadband antennas} such as log–periodic dipole array, butterfly/fat-dipole designs, and inverted-V or crossed-dipole elements~\cite{PierreAuger:2016qzd,Huege:2019snr,VanHaarlem:2013abc,LOPES:2005ipv,Charrier:2018fle,GRAND:2018iaj,Southall:2022yil}.

\vspace{1cm}
\begin{figure*}[t!]
  \centering
  \setlength{\tabcolsep}{3pt}    
  \begin{tabular}{ccc}
    \includegraphics[width=1.0\linewidth]{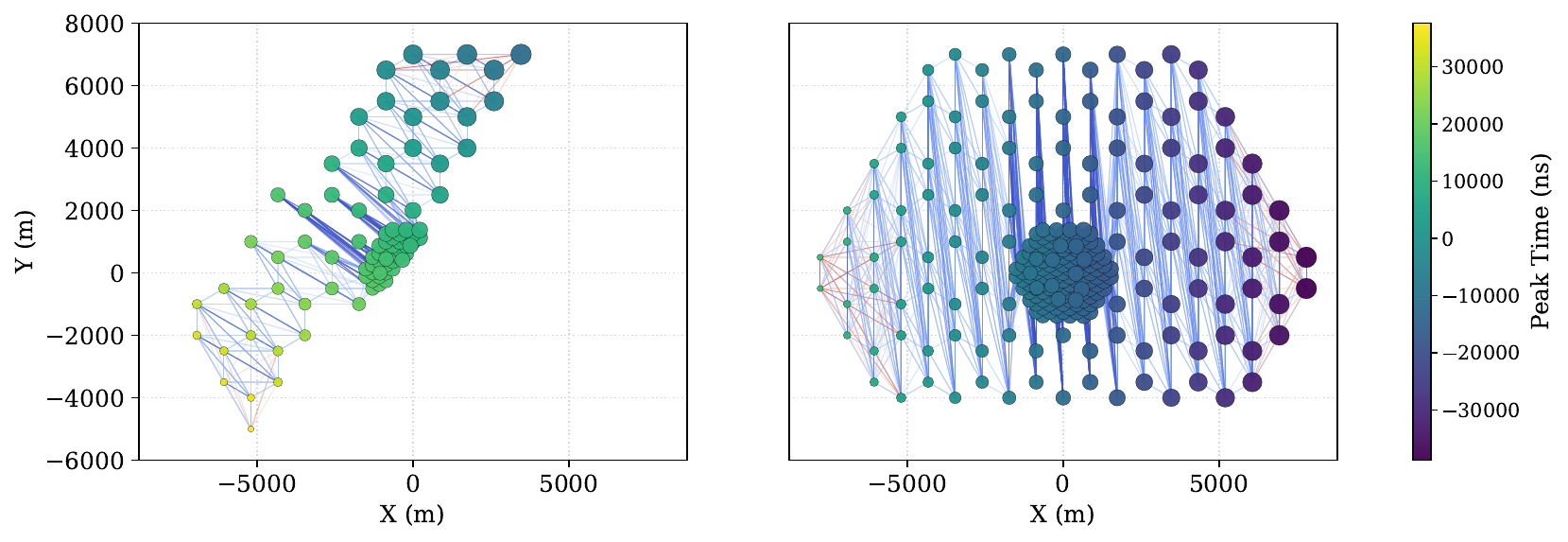}
  \end{tabular}
  \caption{\textbf{Constructed graph representations of two simulated air-shower events using the $k$-nearest-neighbor (kNN) method on the GRANDProto300 geometry.} Triggered antennas are shown at their $(x,y)$ locations; marker color encodes relative peak time (ns) and marker size scales inversely with time to highlight early arrivals. Edges connect \emph{temporally} nearest neighbors: for each node we link to $k=\sqrt{N}$ antennas (clipped to $3 \le k \le 12$) with the smallest $|\Delta t|$, where $N$ is the number of triggered nodes. For illustration purposes, only the highest-weight subset of edges is rendered; thus some spatially close stations may appear unconnected if they are not close in time-of-arrival. This emphasizes causal (wavefront) connectivity rather than Euclidean proximity and can produce edge bundles roughly orthogonal to the arrival direction. All edges are used in training; the sparse rendering is for visualization only.}
  \label{fig:kNN-graph}
\end{figure*}

\subsection{Synthetic Data Generation and Radio-Signal Formation}\label{subsec:Zhaires}

To infer the arrival directions of UHECRs, we utilize detailed Monte Carlo simulations of extensive air showers. These simulations~\cite{GRAND:2024tdn} were developed by the GRAND collaboration to facilitate the design and optimization of detection algorithms for GRANDproto300~\cite{Chiche:2024ohe}, a prototype array comprising $\sim 300$ antennas, presently under construction in China's Gobi Desert.

The simulations were constructed with the \texttt{ZHAireS} code~\footnote{\url{http://aires.fisica.unlp.edu.ar/zhaires/}}, a simulation package built on the \texttt{AIRES}~\cite{Halzen:1990vt, Zas:1991jv} air-shower engine and extended to calculate radio emission in detail. \texttt{ZHAireS}~\cite{Alvarez-Muniz:2012} tracks the full shower development in the atmosphere and computes the radiated electromagnetic fields from charged particle motion based on first principles. Each simulation propagates particles through the atmosphere in small steps, using Li\'{e}nard-Wiechert potentials to calculate the emitted electric field from each track segment. 

This simulator inherently accounts for interference effects among the myriad moving charges and includes realistic propagation conditions, such as the altitude-dependent atmospheric refractive index. As a result, \texttt{ZHAireS} can predict the complete radio pulse waveform at any observer position, in both time and frequency domains. In addition, \texttt{ZHAireS} can simulate the response of a given antenna array to the shower: for each simulated event, it outputs the electric field trace expected at each antenna, from which one can extract observables like signal peak times and electric field (E-field) amplitudes (see, e.g., Ref.~\cite{Sanchez:2025ksf} and references therein). 

We generated time-domain \emph{electric-field} traces at ground observer points with \textsc{ZHAireS} for downward-going air showers initiated by protons and iron nuclei, spanning energies from 0.4 to 4 EeV; no antenna/electronics response is applied. Our simulations span zenith angles $37^{\circ} \le \theta \le 87^{\circ}$ with uniform coverage in azimuth ($0^{\circ}$–$360^{\circ}$) and adopt the GRANDProto300 geomagnetic field and atmospheric model.

To reproduce the timing accuracy of the GRANDproto300 digitizer, we take antenna trigger times directly from the \textit{clean} ZHAireS electric-field traces (no frequency bandpass applied). For each polarization component, we apply a Hilbert transform and use its magnitude to define the envelope. The trigger time for each antenna is taken as the instant when the Hilbert envelope reaches its maximum, using the polarization channel with the largest peak as reference. Antennas whose envelope amplitude falls below $60~\mu\mathrm{V/m}$ are discarded, and only air showers detected by at least five antennas are retained. We add a Gaussian jitter of width $\sigma = 5~\mathrm{ns}$ to each trigger time (matching the GPS synchronization uncertainty measured for GRANDproto300~\cite{GRAND:2024atu}). This yields a clean set of geometric timing measurements while keeping amplitude effects separate (the network is trained only on station coordinates and arrival times ${x,y,z,t}$).

Because we do not apply a hardware bandpass here, this $60~\mu\mathrm{V/m}$ envelope cut should not be interpreted as a realistic trigger threshold tied to a specific 50-200 MHz noise floor; it is used only as a simulation-level quality cut to define which antennas provide timing input. A full treatment of amplitude and noise contributions (including bandpass response and gain calibration) will be addressed in a dedicated amplitude-aware study in follow-up work.

\subsection{Graph Neural Network Direction Estimator}
\label{subsec:GNN}

The direction-reconstruction workflow begins with a graph-convolutional encoder whose latent output is passed to the LtU-ILI inference engine.  We train on synthetic cosmic-ray air-shower events generated with \texttt{ZHAireS}.  Each air shower is cast as a graph whose nodes represent antennas that registered a trigger, annotated with their Cartesian positions $(x,y,z)$ and GPS-synchronised trigger times.  Edges are drawn via a k-nearest-neighbor search in joint space–time, a construction that preserves local information of the radio wavefront.  Figure \ref{fig:kNN-graph} visualises the resulting adjacency structure for two representative events.  Further implementation details—edge feature definition, adaptive choice of $k$, channel dimensionality, optimiser configuration, and checkpointing—are provided in Appendix \ref{appx:A1-event-graph}.

The network architecture comprises multiple graph convolutional layers, each applying local message-passing operations to progressively aggregate spatial-temporal features from neighboring antennas. After several convolutional iterations, node features are pooled globally (mean aggregation) into a single embedding vector representing the entire event. A fully connected multilayer perceptron (MLP) then transforms this global embedding into a three-dimensional correction vector $\Delta\mathbf{k}$. Hyperparameters and the chordal loss function are summarised in Appendix~\ref{appx:A2-GCN-architecture}.

This graph-based approach is particularly advantageous for irregular detector configurations, inherently capturing wavefront structures beyond simplistic geometric approximations. However, purely data-driven GCN predictions, while flexible, may yield non-physical solutions when extrapolating to event geometries or conditions not fully represented in the training dataset.

\begin{table}[t]
\caption{Prior distributions for the shower direction parameters used in the LtU‑ILI training. Each Cartesian component of the unit direction vector is drawn independently from a uniform distribution on the interval $[-1,1]$. }
\centering
\begin{tabular}{lcc}
\toprule
Parameter & Range & Distribution \\
\midrule
$k_x$ & $[-1, 1]$ & Uniform \\
$k_y$ & $[-1, 1]$ & Uniform \\
$k_z$ & $[-1, 1]$ & Uniform \\
\bottomrule
\end{tabular}
\label{tab:priors}
\end{table}

\subsection{Plane-Wave Front Prior as a Physics-Informed Constraint}
\label{subsec:PWF}

To overcome limitations of purely data-driven neural networks, we incorporate a physics-informed prior based on the PWF approximation. This model analytically infers cosmic-ray arrival directions by fitting antenna timing patterns to a planar wavefront perpendicular to the shower propagation axis~\cite{Ferriere:2025}. While neglecting curvature introduces systematic offsets---particularly for near-vertical tracks or impact points close to the array core---the model rigorously preserves causal ordering and global geometry. Our encoder follows the blueprint of Ref. \cite{Ferriere:2025csu}, adapted to the present dataset and training strategy.

The GCN output ($\Delta\mathbf{k}$) is combined with the PWF solution through a \emph{residual gating mechanism} (here ``residual'' refers to predicting a small correction relative to a baseline rather than the full quantity, and ``gating'' refers to a learned scalar that controls how much of that correction is applied). Specifically, the event-level embedding from the GCN and the analytically determined PWF direction $\mathbf{k}_{\text {PWF }}$ are concatenated and processed by a MLP to generate a scalar weighting factor $\alpha \in[0, 1]$. The final corrected direction $\mathbf{k}_{\text {final }}$ is computed as:

\begin{equation}
\mathbf{k}_{\text {final }}=\mathbf{k}_{\text {PWF }}+\alpha\, \Delta\mathbf{k}, \quad \text { with } \quad\left\|\mathbf{k}_{\text {final }}\right\|=1 .
\end{equation}

Numerical values for the gating network (layer sizes, activation) and the unit-vector re-projection step can be found in Appendix~\ref{appx:A} and Tab.~\ref{tab:gcn-pwf-architecture}. This residual gating ensures the network's flexibility in modeling subtle curvature-induced corrections while strictly maintaining physical realism imposed by the analytical model.

\begin{figure}[!t]
    \centering
    \includegraphics[width=\columnwidth]{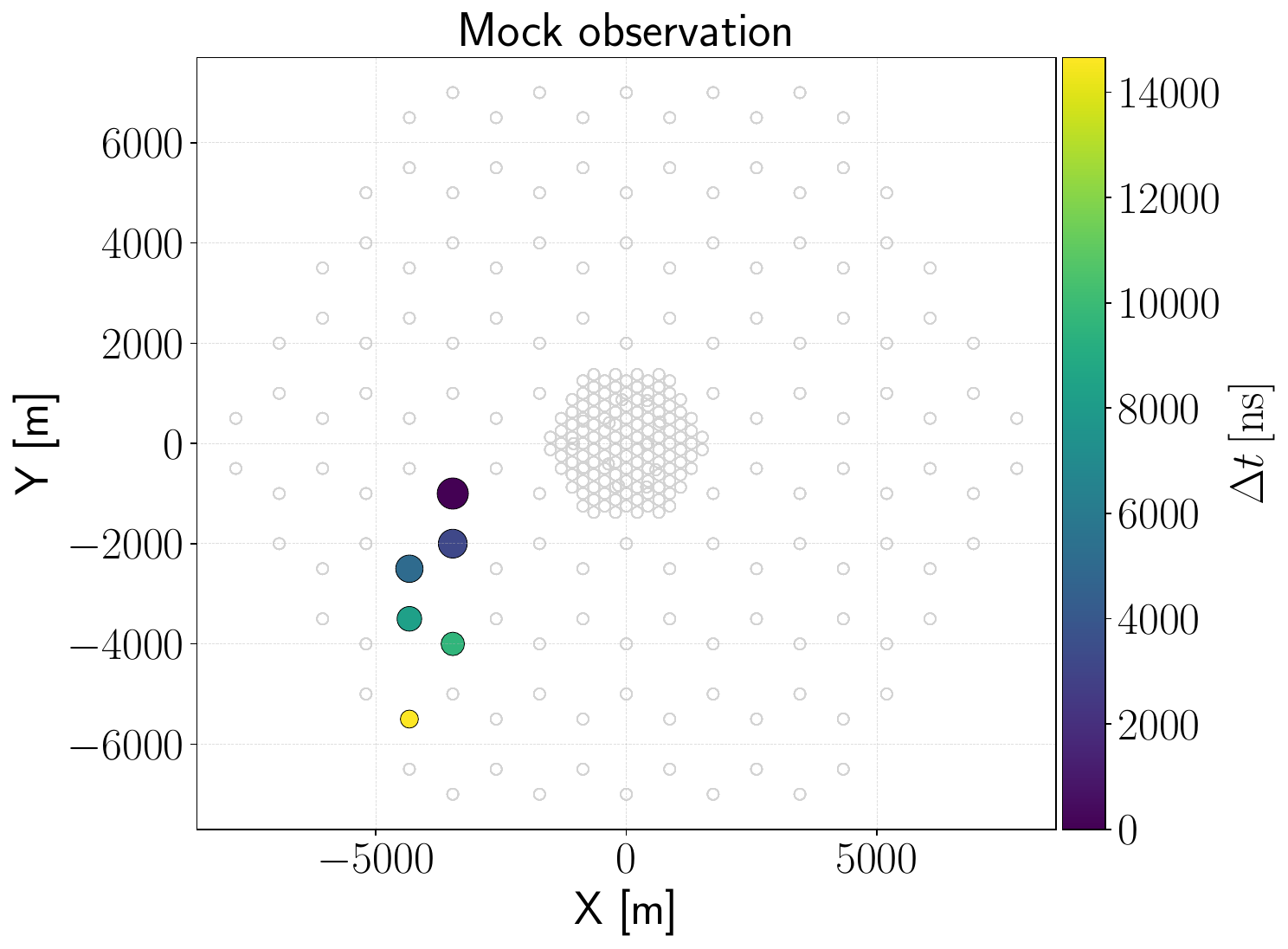}
    \caption{\textbf{Simulated GRANDproto300 ultra-high-energy cosmic ray event drawn from the held-out test set.} Grey open circles represent antenna stations that did not trigger, while colored filled circles denote triggered stations. The size of filled circles decreases with increasing trigger order, and their color encodes the arrival-time delay $(\Delta t)$ relative to the earliest triggered antenna. This synthetic event is extracted from the held-out validation dataset and is used to evaluate the accuracy of our direction reconstruction procedure.}
    \label{fig:hittpattern}
\end{figure}

\subsection{Simulation-based Inference within the \texttt{LtU-ILI} Framework}\label{subsec:Ltu-ILI}

Our inference stage is built on \texttt{LtU-ILI}, which supplies data handling, loss bookkeeping, and ensemble utilities tailored to simulation-based inference. Inside this framework, we deploy an eight-block masked autoregressive flow to represent $p(\theta \mid x)$. Training proceeds with the LtU-ILI optimal-mix schedule: an initial phase with the embedding network frozen, followed by a short gate-only fine-tuning, both regularized by a small entropy term that narrows the density without degrading coverage. Full architecture, hyperparameters, and loss definitions are presented in Appendix~\ref{appx:B-Flow-architecture}.

A training set of approximately 8,200 simulated air-shower events passing predefined quality criteria ( $\geq 5$ antennas triggered above a $60 ~\mu\mathrm{V}$ threshold) was used to train the neural posterior estimators. \texttt{LtU-ILI} manages model optimization by minimizing the negative log-posterior of the reconstructed cosmic-ray direction vectors, employing uniform priors to ensure unbiased generalization. Model training converges typically within 50 epochs on standard GPU hardware. Dataset splits (train/validation/calibration/test) are described in Appendix~\ref{appx:A}.

Specifically, we adopt uniform priors independently applied to each Cartesian component of the direction vector $\mathbf{k}=\left(k_x, k_y, k_z\right)$, as detailed in Table~\ref{tab:priors}. Although this Cartesian prior is not explicitly uniform in zenith and azimuth angles, in practice, this does not significantly influence our results, as the inference process is strongly guided by the PWF model, providing a robust initial direction estimate.

Normalizing flows can be slightly miscalibrated even when accurate in location. We therefore apply a scalar temperature $\tau$ to the learned log-density and renormalize, $p_\tau(\boldsymbol{\theta}\mid\mathbf{x}_{\mathrm{obs}})\propto p(\boldsymbol{\theta}\mid\mathbf{x}_{\mathrm{obs}})^{1/\tau}$. We choose $\tau$ once on a held-out calibration split to match nominal and empirical credibility. Unless stated otherwise, all reported results use this calibrated posterior (details in App.~\ref{appx:B2-temperature-calibration}).

\section{Results}\label{sec:Results}

\begin{figure}[t!]
  \centering
  \begin{tabular}{c}
    \includegraphics[width=\columnwidth]{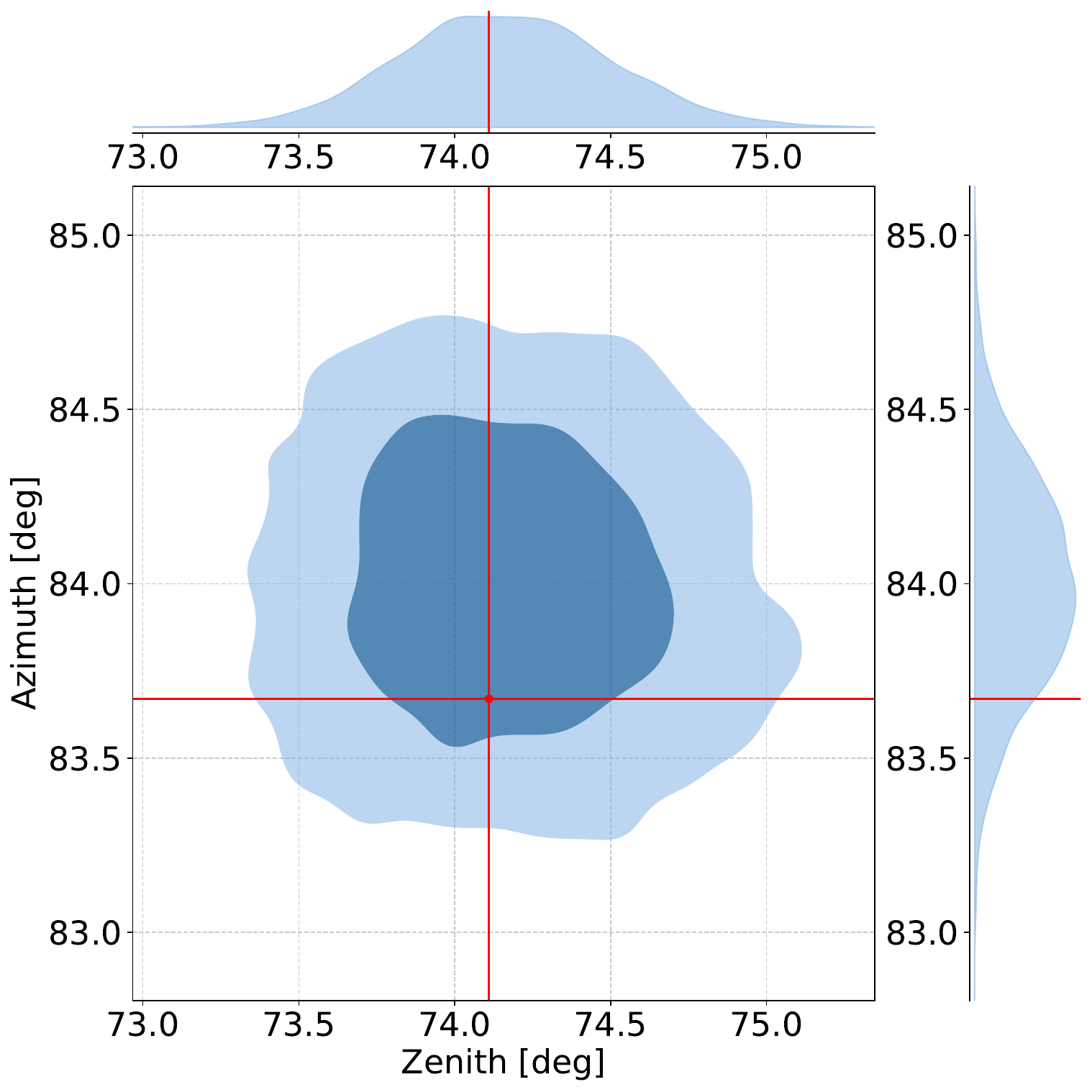} \\
    \includegraphics[width=\columnwidth]{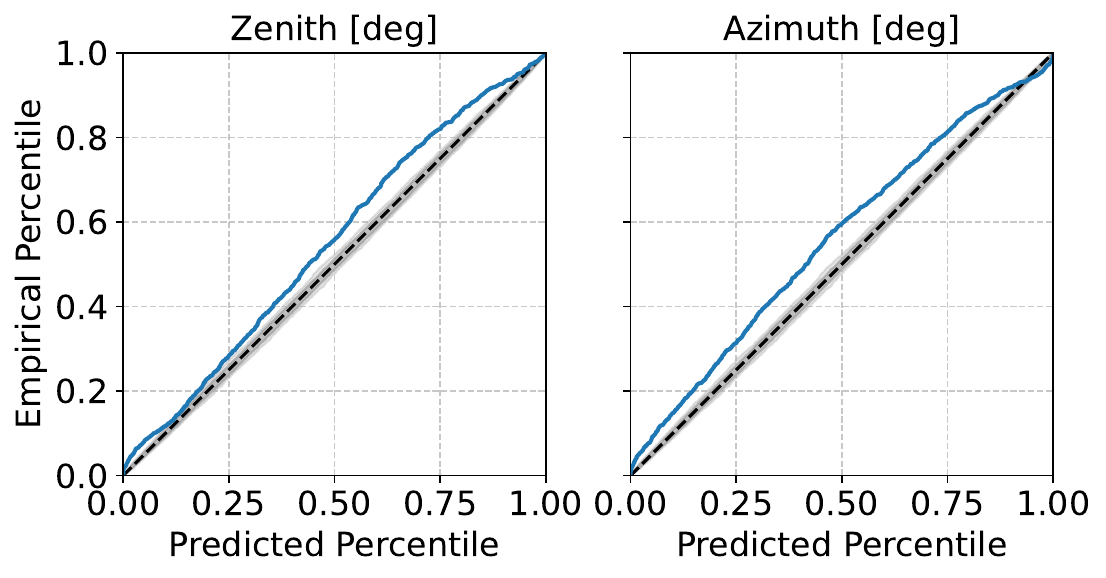}
  \end{tabular}
  \caption{\textbf{Top:} Joint posterior distribution for the mock GRANDproto300-like event shown in Fig.~\ref{fig:hittpattern}. Shaded regions represent the 68\% and 95\% credible intervals for zenith and azimuth, with the true arrival direction marked by a red cross. Corresponding one-dimensional marginal distributions for zenith (top axis) and azimuth (right axis) are also shown. \textbf{Bottom:} Percentile-percentile \textit{(P–P)} plots for zenith (left) and azimuth (right), constructed from $2,048$ test events. Empirical percentiles (blue) lie mostly above the diagonal line and its 95\% bootstrap uncertainty band, indicating mildly conservative posteriors.}
  \label{fig:posteriorstack}
\end{figure}

Following the workflow in Section~\ref{Sec:pipeline}, we first pre-train the embedding network for 50 epochs with the chordal loss (a cosine-complement metric that is smooth in Euclidean space yet honours the unit-sphere geometry, see Appendix~\ref{appx:A3-training-optimization} for further details).

After the encoder converges, we freeze its weights and train the masked-autoregressive flow in two phases, updating only the normalizing-flow head and the gating layers that blend the planar-wavefront seed with the GCN correction (Appendix~\ref{appx:B-Flow-architecture}). We then apply temperature scaling---a post-hoc rescaling $p_T \propto p_\phi^{1 / T}$ that sharpens the posterior when $T<1$---choosing the optimal temperature on a $20 \%$ calibration slice of the validation split.

The network parameters are fitted exclusively on the training data; we tune learning-rate schedules and early-stopping thresholds on the remaining validation set and keep an untouched test set for the final metrics displayed in this section.  Finally, we assessed the calibrated posterior using the coverage and sharpness diagnostics recommended by recent SBI benchmarks~\cite{hermans:2022}, ensuring that the reported credible regions are both compact and statistically well-calibrated.

\begin{figure}[t!]
    \centering
    \includegraphics[width=\linewidth]{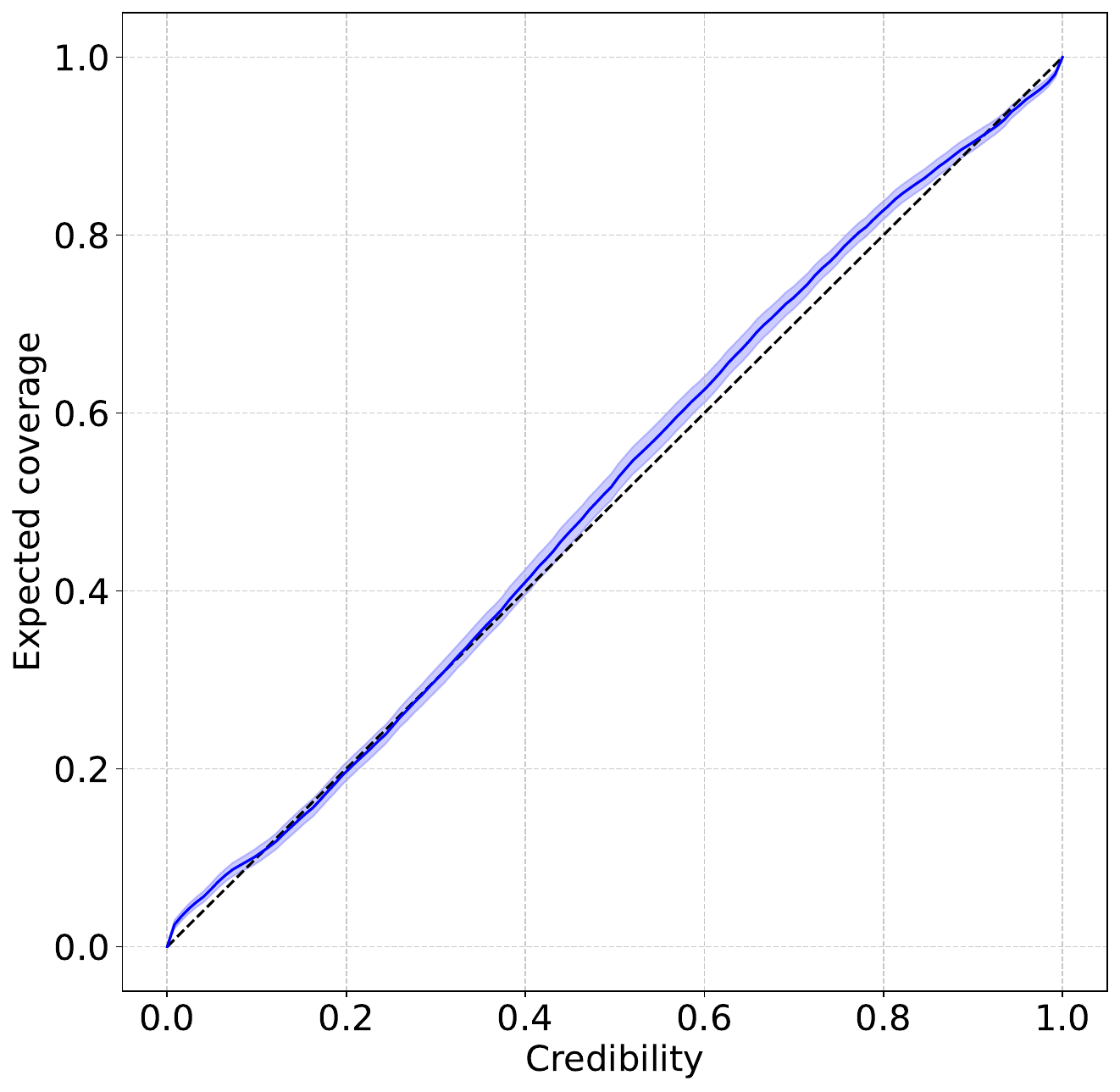}
    \caption{\textbf{TARP (Tests of Accuracy with Random Points) coverage plot for $1,560$ validation events} (see Ref.~\cite{lemos:2023}). The diagnostic was performed directly on the full direction vector $\mathbf{k}$, rather than separately on zenith and azimuth angles. The blue curve represents empirical expected coverage calculated from $2,048$ posterior samples per event, the shaded region indicates a bootstrap-derived $1\,\sigma$ uncertainty band, and the dashed diagonal marks perfect calibration. The observed near-diagonal alignment confirms well-calibrated posteriors, exhibiting mildly conservative uncertainty estimates at credibility levels above $40\%$.
 }
    \label{fig:tarp}
\end{figure}

\begin{figure}[t!]
    \centering
    \includegraphics[width=\linewidth]{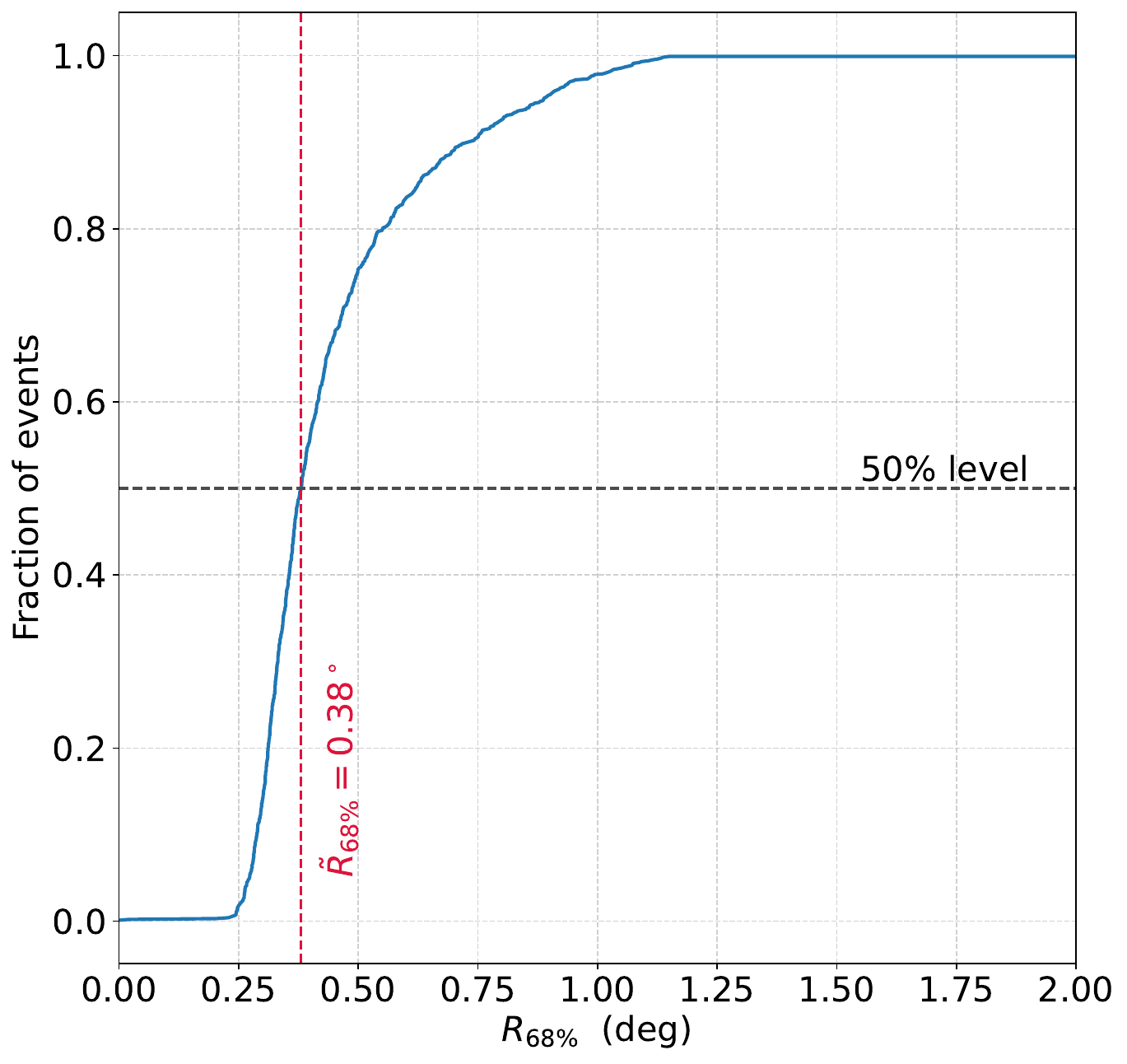}
    \caption{\textbf{Cumulative distribution of the per-event Bayesian containment angle ($R_{68\%}$)} computed at the $68\%$ posterior credible level for an independent validation set consisting of $1,560$ simulated GRAND events. The vertical axis indicates the cumulative fraction of events with containment angles equal to or smaller than the corresponding horizontal axis value. The vertical dashed line marks the median resolution $\tilde{R}_{68}$, characterizing the typical directional precision of the reconstruction pipeline. This distribution encapsulates the complete variability and performance of our inference procedure.}
    \label{fig:cdf}
\end{figure}

\begin{figure*}[t!]
    \centering
    \includegraphics[width=0.49\textwidth]{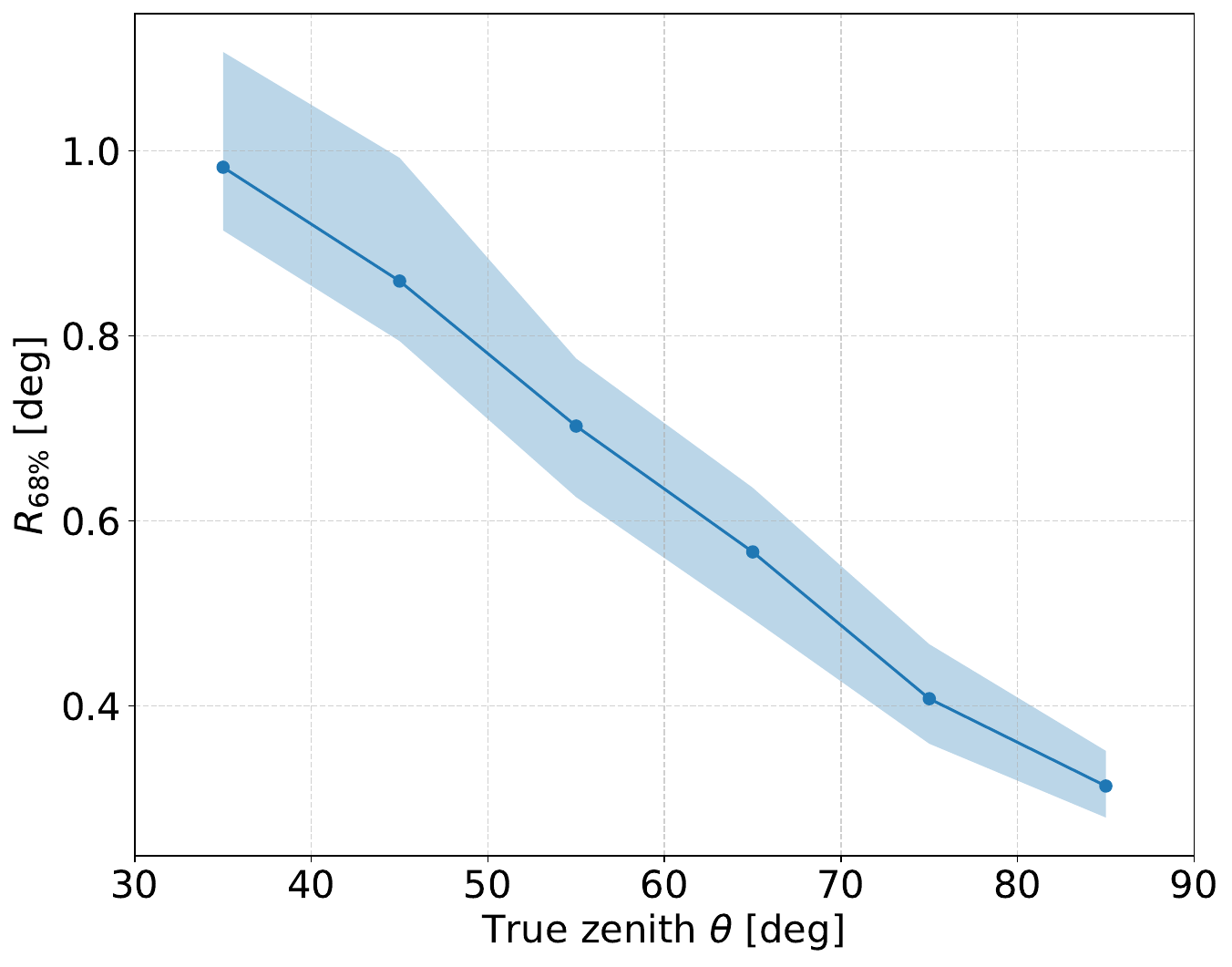}%
    \hfill
    \includegraphics[width=0.49\textwidth]{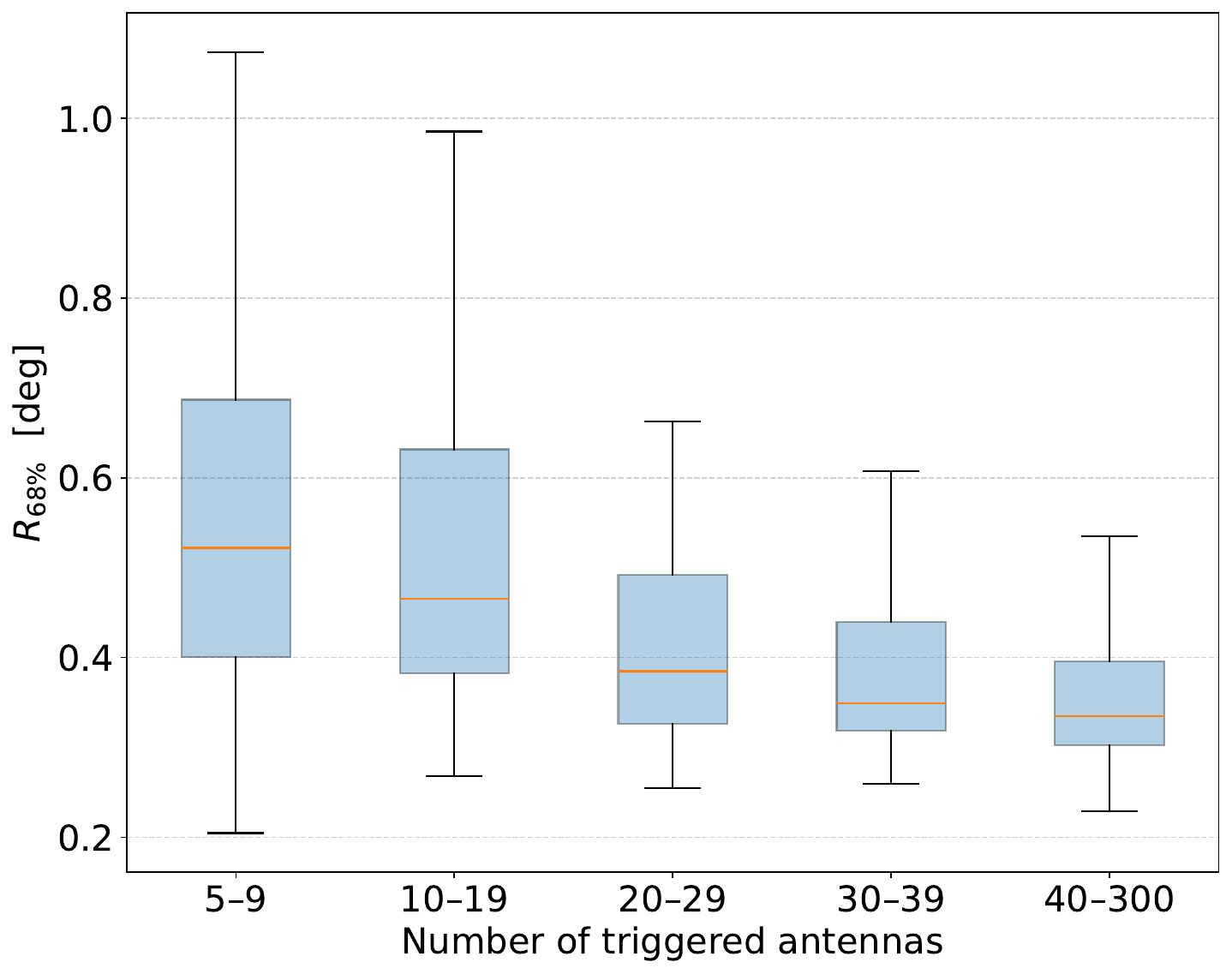}
    \caption{\textbf{Left:} Median Bayesian $68 \%$ containment angle $R_{68\%}$ (circles) as a function of true zenith angle $\theta$, computed from the same test events considered in Fig.~\ref{fig:cdf}. Events are grouped into seven equal-width zenith angle bins, with shaded bands denoting the 16th--84th percentile range of $R_{68\%}$ within each bin. \textbf{Right:} Dependence of the Bayesian $R_{68\%}$ containment angle on triggered-antenna multiplicity, quantified by the number of triggered antennas per event ($N_{\rm{trig}}$). Events are grouped into bins of antennas ( $0-9,10-19$, ..., $40+$). The median values are represented with orange horizontal lines, interquartile ranges with boxes, and the 2.5\%--97.5\% intervals with whiskers. The observed systematic decrease in both median and spread of $R_{68\%}$ with increasing $N_{\rm{trig}}$ indicates enhanced directional precision due to the additional timing constraints, achieving sub $\approx 0.6^{\circ}$ median angular resolution for $N_{\rm{hit}} \geq 30$. }
    \label{fig:angular-resolution}
\end{figure*}

In Fig.~\ref{fig:hittpattern}, we illustrate the antenna trigger pattern of a representative validation event. Open circles denote antenna stations that remained inactive during the mock observation, whereas filled circles indicate the six antennas that triggered signals exceeding the detection threshold. The size of each filled circle decreases according to the trigger order, and their color indicates the relative arrival-time delay $\Delta t$ with respect to the earliest triggered antenna. Importantly, since this event originates from the validation dataset---entirely separate from the test dataset used in hyperparameter tuning---it provides an unbiased benchmark for evaluating the performance of the final physics-informed graph convolutional network integrated with an eight-block Masked Autoregressive Flow.

Figure~\ref{fig:posteriorstack} (top) presents the joint posterior distribution in zenith and azimuth for the same validation event depicted in Fig.~\ref{fig:hittpattern}. The 68\% and 95\% credible intervals contain the true event direction (marked by a red cross), demonstrating sub-degree reconstruction accuracy for the direction of this particular event. Marginal posterior distributions, displayed along the axes, concentrate most of the mass probability around the true values. 

Figure~\ref{fig:posteriorstack} (bottom row) presents percentile–percentile \textit{(P–P)} plots constructed from 2,048 independent validation events, comparing nominal posterior quantiles to their empirical coverage frequencies. The empirical curves closely follow the ideal one-to-one line, though a minor upward deviation at all quantiles suggests mildly conservative credible intervals, indicating slightly broader coverage than strictly required. This slight excess coverage is desirable in astrophysical analyses, where underestimating uncertainty carries a higher cost than modest conservatism \cite{hermans:2022}.

To supplement the \textit{P-P} coverage tests, we subjected our inferred posteriors to the \textit{Tests of Accuracy with Random Points} (TARP) diagnostic introduced in Ref.~\cite{lemos:2023}.  TARP compares the nominal credibility $\alpha$ of many randomly-centered hyper-spherical regions to the empirical coverage probability estimated from posterior samples.  Unlike traditional highest-posterior-density or simulation-based-calibration tests, TARP relies only on posterior samples and is both a necessary and a sufficient condition for calibration. In principle, the identity line is recovered if the approximate posterior equals the true one for every datum.  Consequently, it can reveal subtle pathologies that other diagnostics may miss.

We ran the TARP coverage test in the \texttt{LtU-ILI} framework on the direction vectors $\mathbf{k}$. For each of the $1,560$ held-out events we drew 2,048 posterior samples from the flow ensemble. The expected-coverage curve hugs the one-to-one line across the full credibility range, drifting only slightly above it (mildly conservative) for credibility $\gtrsim 0.4$ (Fig. \ref{fig:tarp}). In particular, the 2,048 posterior samples were drawn independently from the trained MAF posterior for a fixed mock event $\mathbf{x}_{\mathrm{obs}}$ (we sample $z\sim\mathcal{N}(0,I)$ and map $z$ to $\boldsymbol{\theta}=f^{-1}(z;\,\text{embedding}(\mathbf{x}_{\mathrm{obs}}))$) with no re-simulation of timing jitter at inference  (see Ref.~\cite{papamakarios:2018} for MAF sampling details). Our MAF flow model is described in App.~\ref{appx:B-Flow-architecture}.

Overall, these results demonstrate that the pipeline’s posteriors are not merely well-calibrated on average---a necessary but insufficient criterion---but also satisfy the more stringent condition given by the TARP diagnostic. Consequently, our approach reliably provides accurate and rigorously validated uncertainty estimates for UHECR arrival directions across the entire validation dataset.

\subsection*{Angular‐resolution performance of the SBI pipeline}
\label{subsec:angular-resolution}

We measure directional precision with the Bayesian $68 \%$ highest-posterior-density angle, $R_{68\%}$, defined as the radius of the smallest spherical cap that contains $68 \%$ of the posterior mass for a single event. Because $R_{68\%}$ is rotation-invariant and tied directly to posterior mass, it avoids the coordinate artifacts that can distort marginal uncertainties in zenith and azimuth~\cite{Bonifazi:2009}. We compute $R_{68}$ for all 1,560 events in the hold-out validation split, representing each posterior with 2,048 Monte-Carlo samples; this depth keeps the Monte-Carlo noise in the quoted quantiles below $1 \%$.

Figure~\ref{fig:cdf} displays the cumulative distribution function (CDF) of $R_{68\%}$. We find that the median resolution is $
\tilde{R}_{68\%}=0.38^{\circ}$. Only $\simeq 3 \%$ of events exceed $1^{\circ}$, demonstrating that the heavy tail is short. We report medians rather than means because the few worst-reconstructed events unduly influence the latter. The inter-quantile range analogously supersedes the traditional standard deviation for skewed, bounded errors.

\subsubsection*{Dependence on arrival geometry}
\label{subsubsec:arrival-geometry}

We now examine whether the pipeline's angular resolution varies systematically with event geometry. Figure \ref{fig:angular-resolution} (left) reveals a clear zenith trend in the $68 \%$ highest-posterior-density angle $R_{68\%}$. Resolution improves steadily from vertical showers to near horizontal events, with the tightest constraints at the most inclined geometries. The shaded band (median $\pm$ interquartile range) stays below $\approx 1^{\circ}$ across the full zenith span, showing that residual scatter arises mainly from measurement noise rather than zenith-dependent bias. We emphasize that we report median and interquartile range to capture the central tendency and spread without assuming Gaussian errors.

\subsubsection*{Impact of number of triggered antennas}
\label{subsubsec:antenna-multiplicity}

We also investigated how angular resolution depends on the number of antennas triggered by an event. Figure~\ref{fig:angular-resolution} (right) categorizes events according to the number of triggered antennas, $N_{\rm{trig}}$. The median $R_{68\%}$ improves systematically from $1.1^\circ$ in the lowest bin ($5$–$9$ antennas) to $0.5^\circ$ for events with $N_{\mathrm{trig}} \geq 40$, accompanied by a concurrent narrowing of the interquartile range. This monotonic improvement aligns with expectations based on larger statistics, as additional timing measurements progressively constrain the directional reconstruction.

\subsubsection*{Timing-Noise Information Limit}
\label{subsubsec:timing-noise-info-limit}

\begin{figure}[t!]
    \centering
    \includegraphics[width=1.0\linewidth]{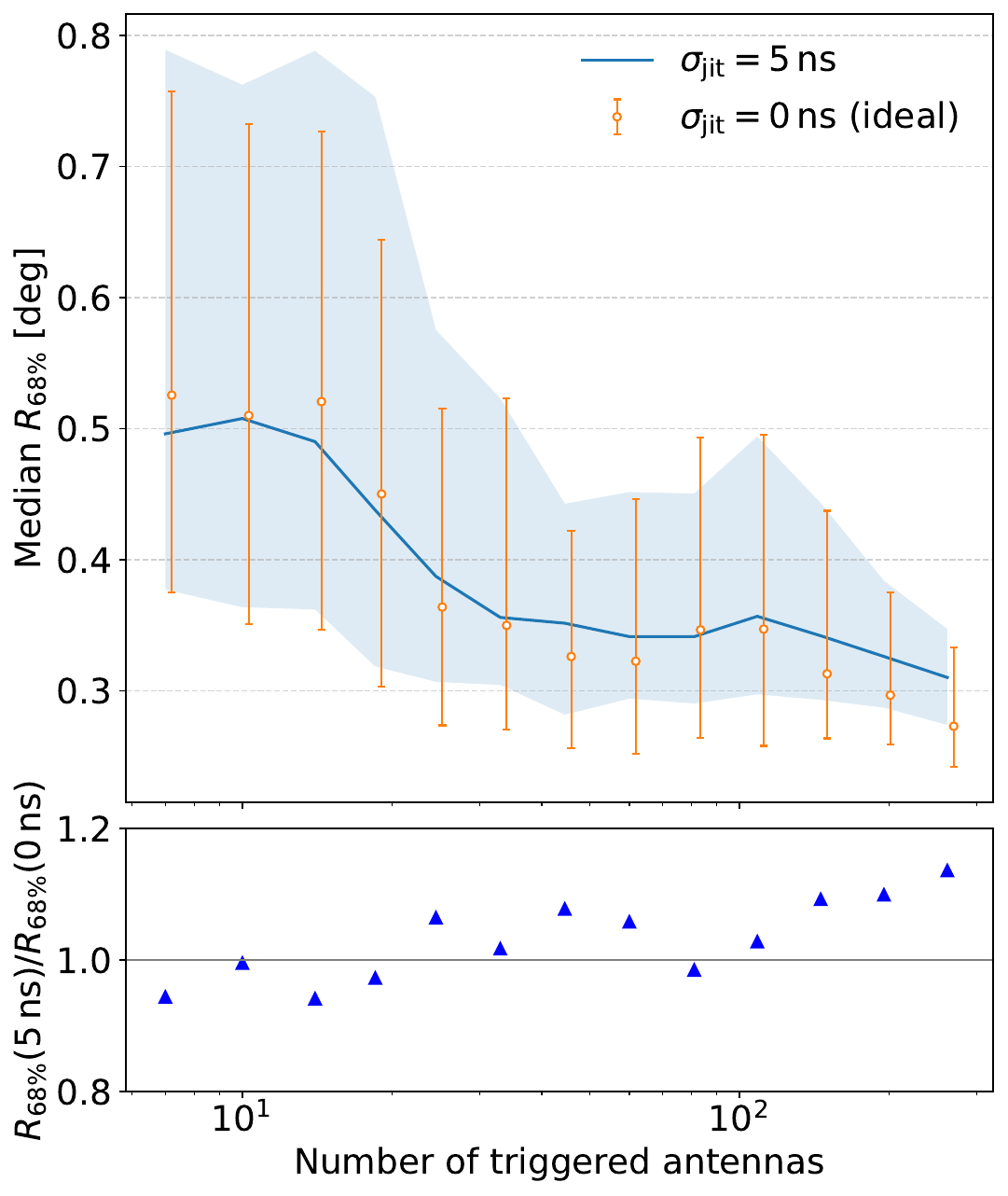}
    \caption{\textbf{Timing-noise information limit.} Median $68 \%$ highest-posterior-density (HPD) angular radius $R_{68\%}$ is shown versus triggered-antenna multiplicity for $5 \mathrm{~ns}$ jitter noise (solid blue, $16-84 \%$ band) and no noise (orange markers, $16-84 \%$ bars). Finer multiplicity bins than in Fig.~\ref{fig:angular-resolution} (right) clarify the high-multiplicity plateau that emerges once $\gtrsim$ 30 antennas are triggered. The residual panel ($R_{68\%}^{5 \mathrm{~ns}} / R_{68\%}^{0 \mathrm{~ns}}$ vs. multiplicity) shows that GPS jitter, not array geometry, sets the ultimate angular-resolution floor.}
    \label{fig:infolimit}
\end{figure}

We find that once an event triggers $\gtrsim 30$ antennas, the median directional error plateaus---it settles near $0.30^{\circ}$ for the baseline jitter noise ($\sigma_{\mathrm{jit}}=5 \mathrm{~ns}$)---yet the spread of the posterior keeps tightening (Fig.~\ref{fig:infolimit}, upper panel). In that regime, each additional station contributes additional statistics, which helps narrow down the 68\%  credible band by roughly a factor of two between 30 and 200 signals, but cannot pull the centroid below the stochastic clock offset. The residual plot in the lower panel of Fig.~\ref{fig:infolimit} confirms that timing noise, and not geometry, governs the angular resolution floor.

Our results suggest that adding more antennas will continue to narrow the posterior, but pushing the centroid error below the $\approx 0.3^{\circ}$ plateau would require sub-nanonsecond timing instrumentation. However, a complementary approach~\cite{Ferriere:2025csu} would be to enrich our embedding network input by supplying per-station electic field amplitudes or energy fluence, which could allow the encoder to down-weight inconsistent arrival-amplitude combinations, thereby trimming the residual bias without requiring a hardware upgrade.

\subsection*{Comparison with analytic UHECR direction reconstruction methods}
\label{subsec:comparison-with-analytic-recon}

\begin{figure}
    \centering
    \includegraphics[width=1.0\linewidth]{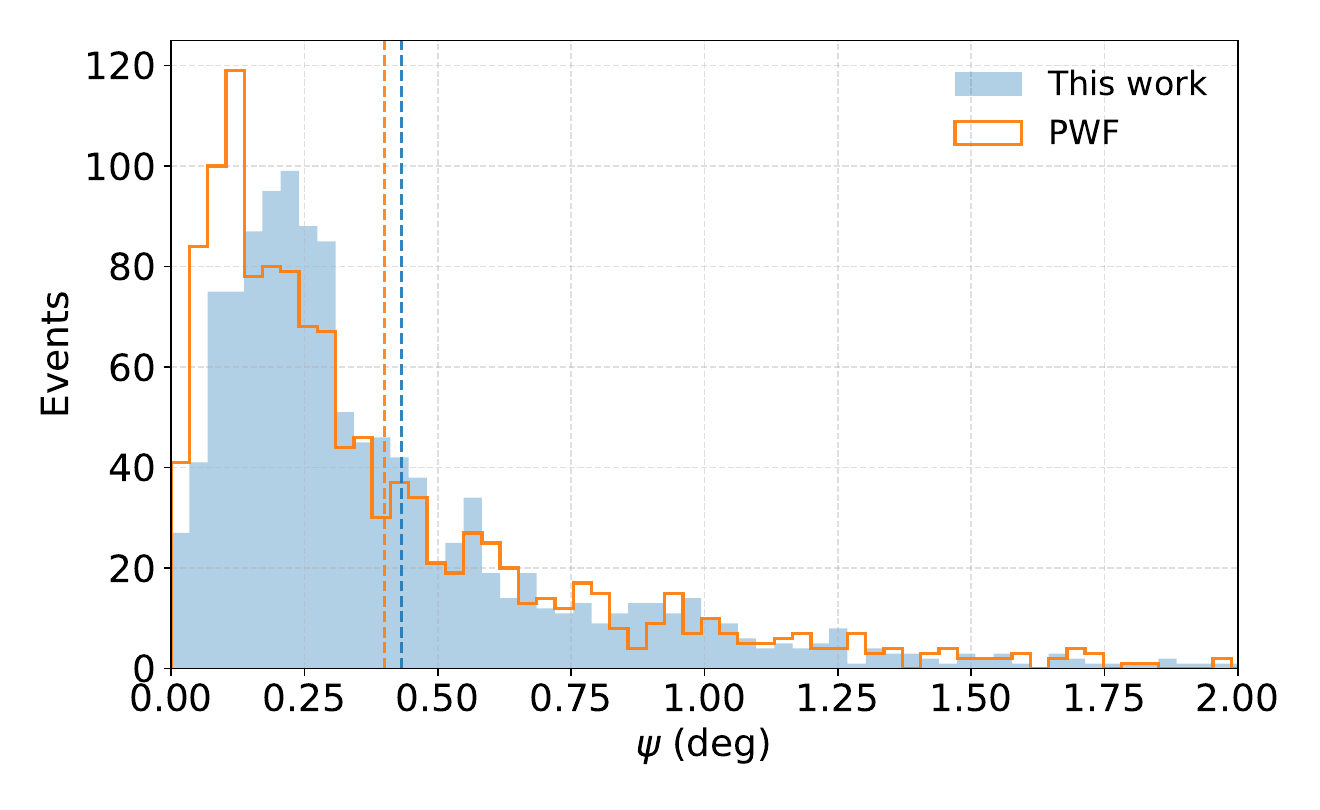}
    \caption{\textbf{Comparison of the point-estimate angular resolution of the PWF-prior fed into the network (outline histogram) and our trained SBI model (filled histogram).} The angular error is computed as $\psi=\arccos(\hat{\boldsymbol k}\cdot\boldsymbol k_{\rm true})$, where $\hat{\boldsymbol k}$ and $\boldsymbol k_{\rm true}$ are the reconstructed and true direction vectors, respectively. The vertical dashed lines show the mean ($\mu_{\rm PWF}=0.40^\circ$, $\mu_{\rm SBI}=0.43^\circ$) for both methods evaluated on the same held-out test events.} 
    \label{fig:comparison-PWF-SBI}
\end{figure}

Analytic direction reconstruction pipelines can reach very small \textit{point-estimate} errors under certain conditions. In particular, for highly inclined showers ($\theta \gtrsim 60^{\circ}$) using timing and amplitudes, the Angular Distribution Function method~\cite{Guelfand:2025goo} reports a mean direction error better than $4^{\prime}\left(\approx 0.07^{\circ}\right)$ on mock observations. Similarly, the PWF method introduced in Ref.~\cite{Ferriere:2025}, which uses trigger times and antenna positions, reports a mean error of approximately $0.2^{\circ}$ for a GPS jitter of $10$ ns. These quoted resolutions were obtained on simulation datasets that differ from the one used here (so absolute values are not directly comparable).

We observe that we do not aim to outperform PWF in point-estimate resolution in this work; our focus is calibrated uncertainty. Nevertheless, it is instructive to compare the direction-reconstruction resolution of the PWF~\cite{Ferriere:2025} and our amortized SBI model on equal footing. For this, we evaluate both models on the same events from the held-out test split. For the PWF, we use the standalone public package (\texttt{PWF\_reconstruction}~\footnote{\url{https://github.com/arsenefer/PWF_reconstruction}}).   

Figure~\ref{fig:comparison-PWF-SBI} shows histograms of the point estimates obtained with both methods. As can be seen, our amortized SBI model has learned to defer to the PWF-prior for virtually all events.  However, our SBI model does learn the event-by-event uncertainty around the PWF prior. We expand on this in the Appendix, where we assess the direction uncertainty calibration by comparing empirical coverage to target credibility.

We note that analytic methods typically quote a single angular ``error'' around a point estimate, which implicitly fixes a parametric uncertainty model. In contrast, with our SBI approach, we learn the full posterior $p(k \mid x)$ non-parametrically from the simulator using an amortized normalizing flow, then audit coverage with simulation-based calibration ($P-P$ and TARP plots). When small global deviations are detected, the learned posterior is temperature-calibrated on a held-out split to align nominal and empirical coverage, after which diagnostics are re-checked. This workflow yields probabilistically meaningful credible regions, which will be relevant objects for multi-messenger follow-ups and source identification.

\section{Summary and Discussion}
\label{sec:discussion}

Interestingly, we found that the pre-trained physics-informed embedding (GCN + PWF), even before it was coupled to the normalizing flow, already delivered sub-degree accuracy: the median absolute angular error for events with a large number of triggered antennas ($\gtrsim 40$) was $\approx 0.3^{\circ}$. When we wrapped this embedding inside the masked autoregressive flow to obtain full posteriors, however, the picture changed. The flow now issued 68\% highest-posterior-density cones with a median half-opening angle of $\approx 1.5^{\circ}$, far wider than the empirical scatter. We then ran various coverage tests, which demonstrated that the learned posteriors were clearly overdispersed, even after exploring many different hyperparameter setups. This observation was the prime motivation for introducing the posterior calibration step explained in Appendix~\ref{appx:B2-temperature-calibration}.

As detailed in the appendix, to realign the model's stated credibility with its empirical performance, we applied a single-parameter temperature rescaling, $p_T(\theta \mid x) \propto p(\theta \mid x)^{1 / T}$. We tuned the sole parameter, $T^{\star}=0.90$, on an exclusive 20\% calibration split, minimizing the absolute gap between nominal and observed highest-posterior-density coverage. Because temperature scaling acts monotonically on log-densities, it preserves the learned ordering of directions while compressing radial uncertainty. When we deployed the calibrated flow on an untouched validation set of $N=1,560$ events, the nominal $68\%$ cones enclosed the true direction in $71 \pm 2 \%$ of cases, comfortably within the confidence band around the target.

Our simulation-calibrated posteriors pass both $P-P$ and TARP diagnostics, indicating that the quoted credible regions are \textit{not} over-confident across the entire validation sample. Because calibration is performed with the same simulator that generated the training data, the pipeline implicitly marginalizes over all nuisance effects present in those simulations (e.g., geomagnetic variation, atmospheric refractivity) without imposing parametric forms.

\subsubsection*{SBI pipeline performance}

Building on the calibrated posterior described above, we evaluated the performance of our SBI pipeline on the full event test set. We found that the temperature-scaled flow now encloses the true arrival direction within its $68\%$ highest-posterior-density cone with a median half-opening angle of $0.38^{\circ}$; only $3\%$ of events exceed an angular resolution of $1^{\circ}$. Because those cones derive directly from the posterior, each event carries a simulation-calibrated uncertainty rather than an error bar inferred from population statistics.

For context, the analytic Angular Distribution Function method~\cite{Decoene:2021ncf, Guelfand:2025goo} reports a sharper $\approx 0.1^{\circ}$ direction resolution under similar jitter noise. This is similar to the resolution found in the case of the plane wavefront reconstruction method discussed in Ref.~\cite{Ferriere:2025}. However, it should be noted that Ref.~\cite{Ferriere:2025} provides event-dependent analytic Gaussian uncertainty ellipses and validates them with simulation ensembles; it does not construct an event-level Bayesian posterior. Our pipeline complements this by delivering calibrated Bayesian posteriors with per-event highest-posterior-density credibility regions.

As shown in Fig.~\ref{fig:angular-resolution}, the angular resolution of our pipeline is most pronounced for inclined showers ($\theta \gtrsim 70^{\circ}$). In that regime, the radio footprint spans hundreds of antennas, and the encoding network captures enough shower information to shrink the median highest-posterior-density radius to $\approx 0.30^{\circ}$. Less inclined events ($\theta<40^{\circ}$) remain limited to $\approx 1.0^{\circ}$, yet all credible cones stay below the $1.2^{\circ}$ coincidence window. Furthermore, once an event triggers more than $30$ antennas, the median error plateaus at $\approx 0.35^{\circ}$; though additional stations help tighten the credible region. This floor traces back to the assumed 5 ns GPS jitter.

\subsubsection*{Limitations of our direction-reconstruction SBI pipeline}

We note several simplifications that currently limit the accuracy of our direction reconstruction. Our calibration accounts only for random clock jitter, neglecting other potential sources of instrumental uncertainties such as frequency-dependent hardware delays. These unmodeled effects can systematically shift reconstructed centroids, especially over the largest baselines. Integrating these systematic offsets explicitly into the simulation pipeline and training the graph neural network to recognize invariant patterns would significantly reduce hidden biases.

Further improvement in angular resolution could be achieved by extending the data inputs beyond trigger times alone. By including per-station amplitudes, polarization, and low-frequency spectral features, we could simultaneously reconstruct shower energy and depth of maximum. Leveraging these additional observables would likely push the centroid uncertainty below the intrinsic timing limit demonstrated here, enhancing both the resolution and physical interpretability of the inferred parameters. We will pursue these enhancements in future work.

\subsubsection*{Generalisability and Future directions}

Our encoder processes a four-dimensional $(x,y,z,t)$ input vector for each antenna, dynamically building a k-nearest-neighbour graph. Though our architecture does not hard-code a specific antenna array layout, adapting our model to other layouts such as  
AugerPrime Radio~\cite{PierreAuger:2016qzd}, or BEACON~\cite{Southall:2022yil}, could be achieved by retraining on the target layouts. Our current simulations include contained events only; performance on uncontained cores or markedly different geometries must be established by retraining on representative simulations and revalidating the posterior calibration. Additionally, since our residual-gated network combines an analytic plane-wave fit with data-driven corrections, other first-order wavefront approximations (e.g., spheroidal or hyperbolic wavefronts, or template libraries) could be integrated without substantial modifications.

Embedding our directional inference within a hierarchical normalizing flow conditioned on shower energy and air shower maximum depth ($X_{\rm max}$) would enable detailed composition analyses and energy-dependent studies. Extending the node-level inputs with electric field amplitudes or polarisation information would leverage amplitude-timing correlations, thus improving reconstruction precision.

\section{Conclusions}\label{sec:conclusion}

We present the first end-to-end simulation-based inference pipeline for UHECR direction reconstruction using radio arrays that achieves a median $68\%$ containment of $0.38^{\circ}$ with correct coverage on realistic GRANDproto300 simulations. While our SBI model can represent non-Gaussian posteriors, the main advantage demonstrated here is not exotic posterior shapes, but rather validated calibration of credible regions under the full simulation procedure used to generate the training data.

Because the posteriors are temperature-calibrated on an independent validation slice, the $68 \%$ and  95\% credible regions are expected to enclose the true arrival direction at the stated frequencies, provided the simulator remains faithful to nature. This non-parametric calibration folds all simulated sources of uncertainty (timing jitter, geomagnetic scatter, atmospheric variation) into a single, robust error budget that can be propagated directly into astrophysical studies.

Looking ahead, integrating amplitude information and extending the flow to joint direction-energy composition space will enable bias-free, fully Bayesian characterization of ultra-high-energy cosmic rays and neutrinos at next-generation radio arrays, positioning simulation-based inference as a keystone methodology for the coming decade of multi-messenger astronomy.

\section*{Data Availability}
\label{sec:data-availability}

The training and inference code used in this work is archived on Zenodo \cite{Macias:sbi-package-2025}
(DOI: \href{https://doi.org/10.5281/zenodo.16895985}{\nolinkurl{10.5281/zenodo.16895985}})
and mirrored at \href{https://github.com/oscar-macias/sbi_uhecr_radio_recon}{GitHub}. The training dataset used to produce the reported results consists of GRAND Collaboration proprietary \texttt{ZHAireS} simulations and cannot be redistributed. To facilitate reuse, the repository includes a lightweight synthetic/mock dataset and configuration files that allow users to run the full pipeline and explore the code. Trained model checkpoints derived from the proprietary data are not publicly released; requests for access to the underlying simulations should be directed to the corresponding author and are subject to GRAND Collaboration data policies.

\section*{Acknowledgements}
We thank Claire Gu\'epin, Kumiko Kotera, and Olivier Martineau-Huynh for insightful discussions during the initial stages of this study. We also thank the GRAND Collaboration for helpful discussions and for supplying GRANDproto300 site parameters and related technical details used in this study. OM acknowledges support from the U.S. National Science Foundation under Grant No. 2418730.


\bibliography{ZReferences}

\appendix

\section{Architecture of the physics-informed embedding network model}
\label{appx:A}
\subsection{Event-Graph Construction}
\label{appx:A1-event-graph}

We encode each detected air-shower event as an undirected graph $G=(\mathrm{Nodes},\mathrm{Links})$. Nodes represent triggered antennas with feature vectors $\mathbf{v}_i=(x_i,y_i,z_i,t_i)$, comprising spatial coordinates and GPS-synchronized trigger times (ns). Links connect each node to its temporally nearest neighbors by minimizing the absolute trigger-time difference $|\Delta t_{ij}|$. Each node is connected to a target of $k=\min\!\bigl(12,\lceil\sqrt{|V|}\rceil\bigr)$ neighbors (we then symmetrize the links to form an undirected graph). 
For each link we store two attributes ($|\Delta t_{ij}|$ [ns] and the Euclidean antenna separation $r_{ij}$ [m]) to retain both timing and geometric context. Because neighbor selection depends only on $|\Delta t_{ij}|$, no arbitrary space–time unit scaling enters the graph construction. The choice of $k$ is a practical sparsification heuristic; exploring alternative neighbor counts and combined space–time metrics is left to future work.

\begin{table*}[t]
\caption{Layer architecture of the physics-informed direction-reconstruction network. ReLU activations and dropout ($p=0.1$) follow each hidden layer. Trainable parameters count weights and biases only.}
\centering
\resizebox{\textwidth}{!}{%
\begin{tabular}{@{}l c r@{}}
\toprule
\textbf{Layer} & \textbf{Operations} & \textbf{Trainable parameters} \\
\midrule
GCNConv-1     & Graph convolution: ReLU $\circ$ GCNConv(4 $\rightarrow$ 128)                   & $4\times128 + 128 = 640$    \\
GCNConv-2     & Graph convolution: ReLU $\circ$ GCNConv(128 $\rightarrow$ 64)                  & $128\times64 + 64 = 8,256$ \\
MeanPool      & Global mean pooling of node embeddings                                      & ---                         \\
FC-1          & MLP: ReLU $\circ$ Linear(64 $\rightarrow$ 64), dropout (0.1)                  & $64\times64 + 64 = 4,160$  \\
FC-2          & MLP: ReLU $\circ$ Linear(64 $\rightarrow$ 64), dropout (0.1)                  & $64\times64 + 64 = 4,160$  \\
$\Delta\mathbf{k}_{\rm GCN}$       & MLP output: Linear(64 $\rightarrow$ 3) to predict correction $\Delta\mathbf{k}_{\rm GCN}$ & $64\times3 + 3 = 195$       \\
Gate-1        & Gating: ReLU $\circ$ Linear(67 $\rightarrow$ 64)                             & $67\times64 + 64 = 4,352$  \\
Gate-2        & Gating: ReLU $\circ$ Linear(64 $\rightarrow$ 32)                             & $64\times32 + 32 = 2,080$  \\
Gate-3        & Gating: Sigmoid $\circ$ Linear(32 $\rightarrow$ 1), clamp $[0.1,0.9]$           & $32\times1 + 1 = 33$        \\ 
Final output  & Normalize($\mathbf{k}_{\mathrm{PWF}} + \alpha \Delta\mathbf{k}$), unit norm &  ---\\
\bottomrule
\end{tabular}%
}
\label{tab:gcn-pwf-architecture}
\end{table*}

\subsection{Graph Convolutional Network Architecture}
\label{appx:A2-GCN-architecture}

We detail the embedding architecture in Table~\ref{tab:gcn-pwf-architecture}. Briefly, two graph convolutional layers (GCNConv)~\cite{kipf:2017} transform the initial four-dimensional node features into a 64-dimensional representation, which is globally mean-pooled into a single event-level embedding. An MLP maps this embedding to a directional correction vector $\Delta \mathrm{k}_{\mathrm{GCN}}$. A parallel gating network takes as input the concatenation of $\Delta \mathbf{k}_{\mathrm{GCN}}$ and the analytic plane-wavefront (PWF) direction, outputting a scalar weight $\alpha$ that blends the corrections. The final corrected direction is normalised to unit length, ensuring physical validity. The full encoder and gating network comprise approximately $2.37 \times 10^4$ trainable parameters. 

\subsection{Training and Optimisation}
\label{appx:A3-training-optimization}
We first pre-train the embedding network for 50 epochs using the \textit{chordal loss}---\textit{i.e.,}the cosine-complement distance on the unit sphere,

$$
\mathcal{L}_{\text {chordal }}=1-\hat{\mathbf{k}} \cdot \mathbf{k}_{\text {true }}, \quad \hat{\mathbf{k}}=\frac{\mathbf{k}_{\mathrm{PWF}}+\alpha \Delta \mathbf{k}_{\mathrm{GCN}}}{\left\|\mathbf{k}_{\mathrm{PWF}}+\alpha \Delta \mathbf{k}_{\mathrm{GCN}}\right\|},
$$

which is a metric widely used in rotation-estimation tasks because it preserves Euclidean smoothness yet respects spherical geometry~\cite{geist:2024}.

After convergence, all the embedding weights are frozen; only the parameters of the normalizing-flow head are updated during the subsequent optimal-mix training of the masked autoregressive flow.

In the pre-training phase, we optimize the encoder using Adam ($\mathrm{lr}=2 \times 10^{-3}$), cosine annealing, AMP acceleration, and gradient clipping with a norm of 5. On an NVIDIA RTX A6000, the training completes in approximately 40 minutes. We select the checkpoint with the lowest median angular error on the $15 \%$ validation split to initialise the subsequent flow training.

\subsection{Data Splits and Reproducibility}
\label{appx:A4-data-split}
We partitioned the dataset of 8,200 simulated events deterministically into 70\% training, 15\% validation, and 15\% testing subsets. We further split the validation subset into calibration (20\%) and hold-out (80\%) samples for posterior temperature calibration. To ensure reproducibility, we initialised all random-number generators (\textsc{numpy}, \textsc{torch}, CUDA) with a fixed seed (42), and enforced deterministic CUDA kernels by setting \textsc{torch.use$\_$deterministic$\_$algorithms(True)}.

\section{Masked autoregressive flow architecture and training}
\label{appx:B-Flow-architecture}
Masked Autoregressive Flows (MAFs) build a complex probability density by chaining invertible, autoregressive transformations whose triangular Jacobians allow closed-form log-determinants~\cite{papamakarios:2018}. This property lets us compute the log-density of simulated parameters during training, while still operating in a likelihood-free setting for the data. MAFs therefore serve as efficient neural density estimators within our simulation-based Bayesian framework, combining expressive power with fast sampling at inference time~\cite{Ho:2024whi}.

We model the conditional posterior $p(\boldsymbol{\theta} \mid x)$ with an eight-block MAF model. Each block applies an affine autoregressive transform whose scale and shift are predicted by a three-layer MADE with 80 hidden units,  providing sufficient flexibility to capture the mild multimodality of the arrival-direction posterior without inflating latency (cf. Fig.~\ref{fig:illustration}). The flow ingests the 64-dimensional event embedding frozen from Appendix~\ref{appx:A2-GCN-architecture}. To improve robustness, we instantiate three independent flows with different random seeds and average their densities with equal weights, forming an ensemble. All hyperparameters follow the reference implementation in our customized version of \texttt{LtU-ILI}.

For a given event $x$, we compute its 64-dimensional embedding $h(x)$ with the frozen encoder (the encoder blends the plane-wavefront seed $\mathbf{k}_{\mathrm{PWF}}$ with the learned residual $\Delta \mathbf{k}$ via a scalar gate $\alpha$ ). To draw one posterior sample, we draw $z \sim \mathcal{N}\left(\mathbf{0}, \mathbf{I}_3\right)$ and apply the inverse autoregressive transformation conditioned on $h(x)$ to obtain $\mathbf{k}=f^{-1}(z \mid h(x))$. We then re-normalise $\mathbf{k}$ to unit length. Repeating this procedure yields i.i.d. samples from $p(\mathbf{k} \mid x)$, which we use to form highest-posterior-densities and to compute coverage.

\subsection{Training of the masked autoregressive flow}
\label{appx:B1-MAF-training}

We train the flow with a customized version of \texttt{LtU-ILI}’s \texttt{Lampe} runner, modified for a two-phase schedule:

\begin{itemize}

\item \textbf{Phase 1:} We freeze the entire embedding and minimise

$$
\mathcal{L}_1=-\left(1+\lambda_H\right)\left\langle\log p_\phi(\boldsymbol{\theta} \mid x)\right\rangle
$$

equivalent to the negative log-likelihood plus an entropy penalty of weight $\lambda_H=10^{-3}$. In practice, this small $\lambda_H$ slightly down-weights broad, high-entropy solutions, encouraging the flow to produce sharper posteriors without sacrificing likelihood fit. We found that 40 epochs at a learning rate of $2 \times$ $10^{-4}$ suffice for convergence under this combined objective.

\item \textbf{Phase 2:} We unfreeze only the gating layers that blend the PWF seed with the GCN correction; the deeper $\Delta \mathbf{k}$ network remains fixed.
These $2.1 \times 10^3$ parameters are fine-tuned for 25 epochs with a learning rate of $10^{-4}$.
    
\end{itemize}

Both phases use Adam with a batch size of 64 and the \texttt{Lampe} gradient sanitisation utilities. Training our MAF model on an RTX A6000 requires $\sim 25 \mathrm{~min}$.

\begin{figure}[t!]
    \centering
    \includegraphics[width=1.0\linewidth]{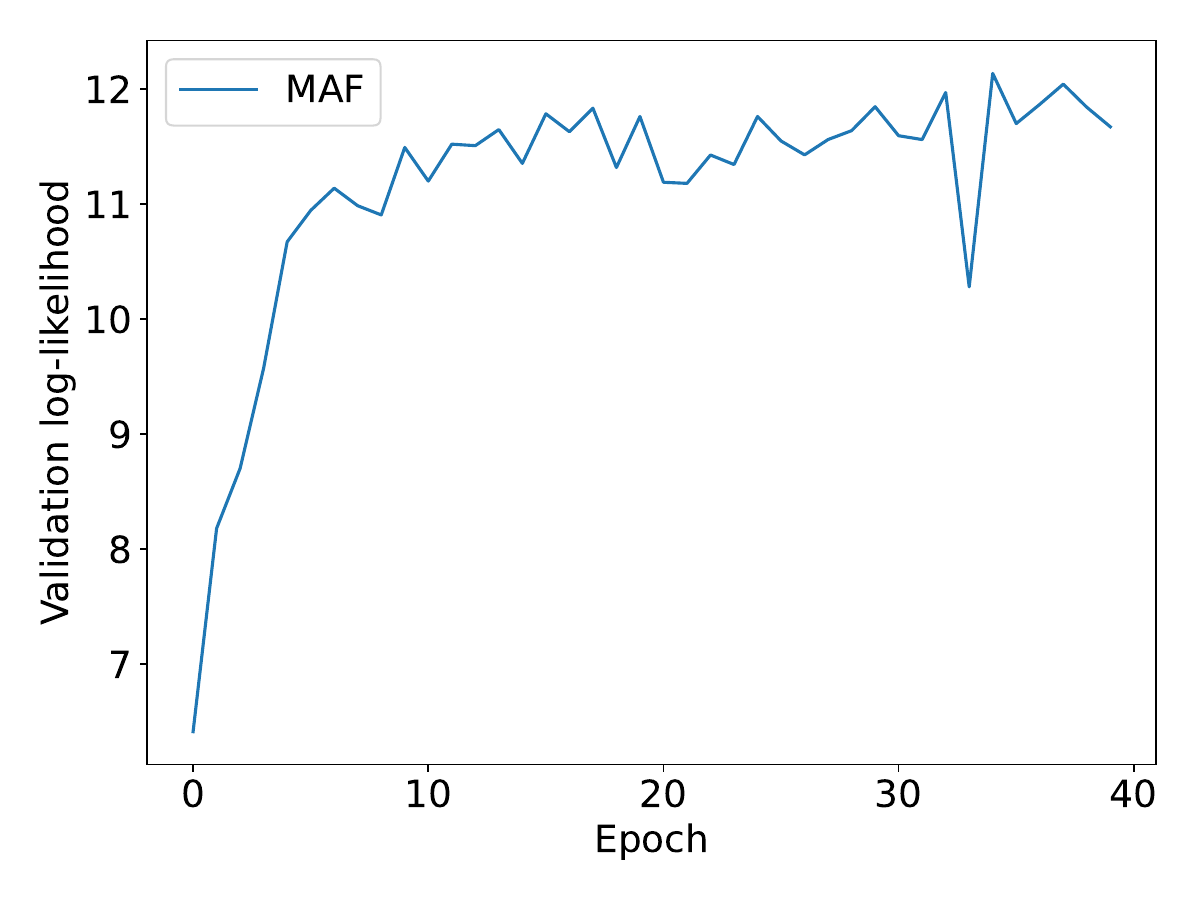}
    \caption{\textbf{Learning curve for the masked autoregressive flow (MAF).} Validation log-likelihood on the held-out dataset versus epochs. The log-likelihood increases steadily during Phase 1 (log-likelihood with a weak entropy penalty) and continues to improve after Phase 2 begins (only the gating layers are unfrozen), with no late-time degradation on validation. This monotonic trend indicates the conditional posterior becomes more faithful on unseen events.}
    \label{fig:loglike}
\end{figure}

Figure~\ref{fig:loglike} shows the learning curve for the masked autoregressive flow. Speficically, the validation log-likelihood on a held-out split as a function of epochs. During Phase 1, the curve rises steadily under the log-likelihood objective with a weak entropy penalty ( $\lambda_H = 10^{-3}$), indicating that the conditional density over directions becomes more faithful to the data. After switching to Phase 2 (only the gating layers that blend the PWF seed with the GCN correction are unfrozen) the log-likelihood continues to improve at a smaller rate without degradation on the validation set (no late-time drop). This behavior demonstrates learning on unseen events and supports that the trained flow captures informative structure in the data rather than overfitting. When training multiple flows, each trace follows the same trend, which indicates the effect is robust to initialization.

\subsection{Temperature calibration of the learned posterior}
\label{appx:B2-temperature-calibration}

We apply temperature scaling to sharpen the learned posterior, following the approach introduced in Ref.~\cite{wenzel:2020}. Starting from the trained flow density $p_\phi(\boldsymbol{\theta} \mid x)$, we define a tempered posterior:

$$
p_T(\boldsymbol{\theta} \mid x)=\frac{p_\phi(\boldsymbol{\theta} \mid x)^{1 / T}}{Z_T(x)}, \quad Z_T(x)=\int p_\phi(\boldsymbol{\theta} \mid x)^{1 / T} d \boldsymbol{\theta}
$$

where $T>0$ is the temperature and $Z_T$ ensures normalisation. For $T<1$, the posterior concentrates around high-density regions, reducing credible intervals, whereas $T>1$ broadens the posterior.

We optimize $T$ by grid-search over the calibration subset ($20 \%$ of validation data), selecting the value that brings empirical $68 \%$ highest-posterior-density  (HPD) coverage closest to the nominal level. We use the smallest-spherical-cap algorithm for HPD evaluation. The optimal temperature, $T^*=0.90$, yields empirical coverage of $71 \pm 2 \%$ on the independent hold-out sample-slightly conservative but well within statistical tolerance. This post-hoc temperature adjustment aligns the Bayesian output with frequentist coverage diagnostics while maintaining the posterior's mathematical coherence and proper normalization via $Z_T(x)$. 

Furthermore, our simulation-based calibration diagnostics ($P-P$ plots and TARP curves) confirmed that the posterior had been sharpened---\textit{not distorted}. In short, this lightweight, transparent correction converts a slightly conservative normalizing flow into a calibrated Bayesian pipeline that retains sub-degree accuracy while delivering trustworthy uncertainty estimates.

\subsection{Comparison of the Direction reconstruction uncertainty between the PWF method in our SBI method}
\label{appdx:comparison-pwf-sbi}

\begin{figure*}[t!]
    \centering
    \includegraphics[width=1.0\linewidth]{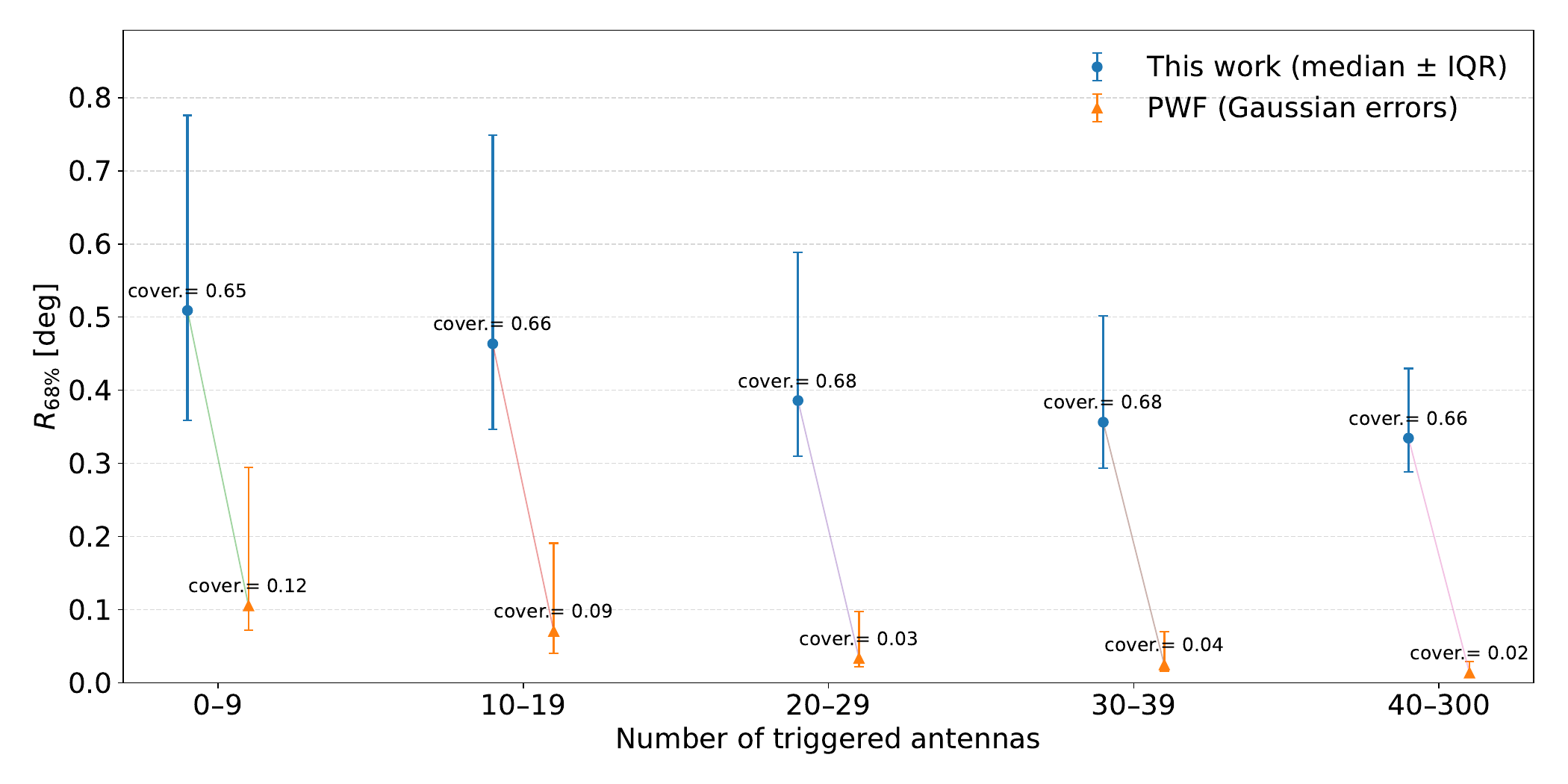}
    \caption{\textbf{Comparison of the angular resolution and coverage for the standalone PWF method and our amortized SBI model.}
The blue points represent  the $R_{68\%}$ radius at $\alpha=0.68$ (median with Interquartile Range (IQR) whiskers) and the orange points show the PWF Gaussian equal-area radius (derived using the standalone PWF package). Labels report the empirical coverage for each method. At a fixed number of triggered antennas, our posteriors achieve convervative angular resolution at near-nominal coverage ($\approx 0.65$ to $0.68$). In contrast, the PWF method achieves much better angular resolution at the cost of under-covering strongly ($\approx 0.02$ to $0.12$). This indicates that the analytic PWF method produces over-confident uncertainties under the planar-wave assumption on realistic simulations.}
    \label{fig:pwfcomparison}
\end{figure*}

We perform a comparison of the direction-reconstruction uncertainty and coverage for the  standalone PWF~\cite{Ferriere:2025} method and our amortized SBI model on the same held-out test events.

For the PWF method, we compute the angular covariance of the PWF solution for each event as~\cite{Ferriere:2025};
\begin{equation}\label{eq:covariance}
\Sigma_{\theta,\phi} \;=\; \mathrm{cov\_matrix}\!\left(\hat\theta,\hat\phi,\{\mathbf x_i\},\sigma_t\right),
\end{equation}
where $(\hat\theta,\hat\phi)$ are the PWF solution for zenith and azimuth, $\{\mathbf x_i\}$ are antenna positions, and $\sigma_t$ is the GPS jitter (here $\sigma_t = 5~\mathrm{ns}$). This yields the standard small-error Gaussian approximation around the best-fit direction, which allows to compute the PWF confidence regions.

We assess \emph{coverage} for the PWF method with a frequentist prescription. Specifically, for each event, we (i) build the PWF error ellipse at a target credibility level $\alpha$ using $\Sigma_{\theta,\phi}$ mapped to the local tangent plane at the PWF solution, (ii) map the true direction into that tangent plane by the shortest great-circle displacement from the PWF estimate, and (iii) record whether this displacement lies inside the ellipse. Aggregating over events (and stratifying by triggered-antenna bins), the fraction inside is the \textit{empirical coverage} at level $\alpha$. If the PWF assumptions held exactly (planar wavefront and correctly specified Gaussian GPS noise), this fraction would match $\alpha$; when it falls short, the ellipses are over-confident. This is what we observe on our test set.

Figure~\ref{fig:pwfcomparison} presents a bin-by-bin comparison at $\alpha=0.68$ between our credible regions and the PWF Gaussian intervals evaluated on the same held-out test events. As can be seen, our $R_{68\%}$ medians decrease with the number of triggered antennas, and empirical coverage remains close to nominal  (approximately $0.65$ to $0.68$) in all bins. This indicates that the amortized SBI model is learning an error structure that matches the held-out data (the intervals are larger than the analytic ones, but they achieve the intended frequency). 

In contrast, PWF Gaussian intervals are much smaller at the same nominal level, yet they under-cover strongly (approximately $0.02$ to $0.12$). This behavior is consistent with a timing-only planar-wave model that ignores additional sources of variability present in realistic radio showers (e.g., wavefront curvature and small modeling mismatches). The point-estimate quality of PWF remains competitive in this regime (particularly at higher number of triggered antennas), but its uncertainty model is over-confident on our simulations. This comparison therefore separates the two results clearly; the PWF method provides a strong prior for direction, whereas our SBI posterior supplies calibrated, per-event uncertainty that maintains nominal coverage on unseen data.

We note that we did not apply any additional calibration (or timing-variance inflation) to the PWF–Gaussian intervals in Fig.~\ref{fig:pwfcomparison}.  That is, they correspond to a ``$\sigma_t$-only'' uncertainty model. Ref.~\cite{Ferriere:2025} explicitly discusses that for realistic ZHAireS simulations the wavefront is not perfectly planar and treats this model discrepancy as an additional timing-variance term (estimated from ZHAireS–PWF time residuals) that can be added in quadrature to $\sigma_t$ to form an effective $\sigma_{\rm tot}$ for interval calibration.

\end{document}